\documentstyle[12pt,psfig]{article} 
\makeatletter \def\@cite#1#2{{[{#1}]\if@tempswa\typeout {IJCGA
warning: optional citation argument ignored: `#2'} \fi}}

\newcount\@tempcntc
\def\@citex[#1]#2{\if@filesw\immediate\write\@auxout{\string\citation{#2}}\fi
  \@tempcnta\z@\@tempcntb\m@ne\def\@citea{}\@cite{\@for\@citeb:=#2\do
    {\@ifundefined
       {b@\@citeb}{\@citeo\@tempcntb\m@ne\@citea\def\@citea{,}{\bf ?}\@warning
       {Citation `\@citeb' on page \thepage \space undefined}}%
    {\setbox\z@\hbox{\global\@tempcntc0\csname b@\@citeb\endcsname\relax}%
     \ifnum\@tempcntc=\z@ \@citeo\@tempcntb\m@ne
       \@citea\def\@citea{,}\hbox{\csname b@\@citeb\endcsname}%
     \else
      \advance\@tempcntb\@ne
      \ifnum\@tempcntb=\@tempcntc
      \else\advance\@tempcntb\m@ne\@citeo
      \@tempcnta\@tempcntc\@tempcntb\@tempcntc\fi\fi}}\@citeo}{#1}}
\def\@citeo{\ifnum\@tempcnta>\@tempcntb\else\@citea\def\@citea{,}%
  \ifnum\@tempcnta=\@tempcntb\the\@tempcnta\else
   {\advance\@tempcnta\@ne\ifnum\@tempcnta=\@tempcntb \else 
\def\@citea{--}\fi
    \advance\@tempcnta\m@ne\the\@tempcnta\@citea\the\@tempcntb}\fi\fi}
\makeatother

\def\boxit#1{\leavevmode\thinspace\hbox{\vrule\vtop{\vbox{\hrule%
        \vskip3pt\kern1pt\hbox{\vphantom{\bf/}\thinspace\thinspace%
        {\bf#1}\thinspace\thinspace}}\kern1pt\vskip3pt\hrule}\vrule}%
        \thinspace}
\def\Boxit#1{\noindent\vbox{\hrule\hbox{\vrule\kern3pt\vbox{
\advance\hsize-7pt\vskip-\parskip\kern3pt\bf#1 \hbox{\vrule height0pt
depth\dp\strutbox width0pt} \kern3pt}\kern3pt\vrule}\hrule}}

\newcommand{\Hh}{\lower1.2ex\hbox{$\stackrel{\textstyle
H}{\footnotesize\sim}$}}
\newcommand{\Hho}{\lower1.2ex\hbox{$\stackrel{\textstyle
H_1}{\footnotesize\sim}$}}
\newcommand{\Hhw}{\lower1.2ex\hbox{$\stackrel{\textstyle
H_2}{\footnotesize\sim}$}}
\newcommand{\h}{\lower1.2ex\hbox{$\stackrel{\textstyle
h}{\footnotesize\sim}$}}

\newcommand{\gsim}{\lower.7ex\hbox{$\;\stackrel{\textstyle>}{\sim}\;$}}
\newcommand{\lsim}{\lower.7ex\hbox{$\;\stackrel{\textstyle<}{\sim}\;$}}
\newcommand{\be}{\begin{equation}} \newcommand{\ee}{\end{equation}}
\newcommand{\beq}{\begin{equation}} \newcommand{\eeq}{\end{equation}}
\newcommand{\bea}{\begin{eqnarray}} \newcommand{\eea}{\end{eqnarray}}

\def\simlt{\stackrel{<}{{}_\sim}} \def\simgt{\stackrel{>}{{}_\sim}}

\def\baselinestretch{1}
\begin{document}
\catcode`@=11 \newtoks\@stequation
\def\subequations{\refstepcounter{equation}%
\edef\@savedequation{\the\c@equation}%
\@stequation=\expandafter{\theequation}
\edef\@savedtheequation{\the\@stequation}
\edef\oldtheequation{\theequation}
\def\theequation{\oldtheequation\alph{equation}}}
\def\endsubequations{\setcounter{equation}{\@savedequation}%
\@stequation=\expandafter{\@savedtheequation}%
\edef\theequation{\the\@stequation}\global\@ignoretrue

\noindent} \catcode`@=12
\begin{titlepage}

\title{{\bf Expectations for LHC from Naturalness:\\
 Modified vs. SM Higgs Sector}} 
\vskip3in \author{   
{\bf J.A. Casas},
{\bf
J.R. Espinosa}
and  {\bf
I. Hidalgo\footnote{\baselineskip=16pt E-mail addresses: {\tt
alberto.casas@uam.es, jose.espinosa@uam.es, irene.hidalgo@uam.es}}}
\hspace{3cm}\\
{\small IFT-UAM/CSIC, 28049 Madrid, Spain}}  
\date{}  \maketitle  \def\baselinestretch{1.15}
\begin{abstract}
\noindent
Common lore has it that naturalness of electroweak breaking  in
the SM requires new physics (NP) at $\Lambda\simlt 2-3$ TeV, hopefully
within the reach of LHC. Moreover the Higgs should be light
($m_h\simlt 219$ GeV) to pass electroweak precision tests
(EWPT). However one should be prepared for ``unexpected'' (although
admittedly unpleasant) results at LHC,  i.e. no NP and/or a heavy
Higgs. We revisit recent attempts to accommodate this by modifying the
SM Higgs sector (using 2-Higgs-doublet models). We find that
these models do not improve the naturalness of the SM, and so they do
not change the expectations of observing NP at LHC. We also stress that
a heavy SM Higgs would not be evidence in favour of a modified Higgs 
sector, provided certain higher order operators influence EWPT. 
On the other hand, we show that  NP can escape LHC detection without a
naturalness price, and with the pure SM as the effective
theory valid at LHC energies, simply if the cut-off for top loops is slightly 
lower than for Higgs loops. 
\end{abstract}

\thispagestyle{empty}
\vspace*{1.cm} \leftline{July 2006} \leftline{}

\vskip-17cm \rightline{IFT-UAM/CSIC-06-36} 
\rightline{hep-ph/0607279} \vskip3in

\end{titlepage}
\setcounter{footnote}{0} \setcounter{page}{1}
\newpage
\baselineskip=20pt

\noindent

\section{Introduction}

It is commonly assumed that the Hierarchy Problem of the Standard
Model (SM) indicates the existence of New Physics (NP) beyond the SM at
a scale  $\Lambda\simlt$ few TeV, hopefully at the reach
of LHC. The argument is well known and goes as follows:

In the SM (taken as the effective theory
valid below $\Lambda$) the mass parameter $m^2$ in the Higgs
potential
\be
\label{VSM}
V= {1\over 2}m^2 h^2 + {1\over 4}\lambda h^4\ ,
\ee
receives dangerous quadratically-divergent
contributions \cite{veltman}. At one-loop, and using a momentum cut-off 
regularization,
\be
\label{quadrdiv0}
\delta_{\rm q} m^2 = {3\over 64\pi^2}\Lambda^2(3g^2 + g'^2 
+ 8\lambda - 8\lambda_t^2)\ ,
\ee
where $g, g', \lambda$ and $\lambda_t$ are the $SU(2)\times U(1)_Y$
gauge couplings, the Higgs quartic coupling and the top Yukawa
coupling, respectively.  The requirement of no fine-tuning between the
above contribution and the tree-level value of $m^2$ sets an upper
bound on $\Lambda$. E.g. for a Higgs mass $m_h = 115-200$ GeV,
\be
\label{quadrft}
\left|{\delta_{\rm q} m^2 \over m^2}\right|
\leq 10\ \Rightarrow \ \Lambda  \simlt 2-3\ {\rm TeV}  \ ,
\ee
where we have implicitly used $m_h^2=2\lambda v^2$
and $v^2=-m^2/\lambda$ (with $v=246$ GeV).
Obviously, if one is stricter about the largest acceptable size 
of $\delta_{\rm q} m^2$,
then $\Lambda^2$ decreases accordingly. E.g.
requiring $|\delta_{\rm q} m^2| \leq |m^2|$ implies 
$\Lambda  \simlt 1\ {\rm TeV}$, well inside LHC reach.
These numbers are somewhat modified when one considers higher order
corrections to eq.~(\ref{quadrdiv0}) (see sect.~2 and 
refs.~\cite{EJ,KM,CEHI}).

There are several reasons to consider possible departures from
this simple SM scenario. First, the above upper bound on $\Lambda$ 
is generically in some tension with the experimental lower
bounds on the suppression scale of higher order operators, derived
from electroweak precision tests (EWPT) \cite{ewfits}, which
typically require  $\Lambda\simgt$ 10 TeV. This is  the so-called
Little Hierarchy problem. It should be kept in mind, however, that if
the NP at $\Lambda\simeq$ few TeV is ``clever enough'', it can still be 
consistent with EWPT. This is the case of Supersymmetric (SUSY) and
Little Higgs models, although these scenarios have their own
fine-tuning problems \cite{CEHI,CEHII}. Second, one should be prepared to
interpret  the possible (though admittedly unpleasant) situation in which
no NP is found at LHC, apart from the Higgs, in spite of the previous 
naturalness arguments based on the simple SM Higgs sector. 
Finally, it could happen
that the Higgs found at LHC is beyond the range consistent with EWPT
($m_h\simlt 186-219$ GeV \cite{EWPTmh}), pointing out again to some 
departure from the ordinary SM Higgs sector.

One of the simplest modifications of the SM Higgs sector one can think of 
is the two Higgs doublet model (2HDM). In a series of recent papers, 
Barbieri et al. \cite{BGH,BH,BHR}, have examined the capabilities of such 
scenario (in different settings) to address the previous questions. Their 
conclusions are that suitable 2HDMs might raise the scale of new physics 
above the LHC reach (keeping naturalness under control) with a light or 
heavy Higgs (depending on the model), in a way consistent with EWPT. Thus 
these models are claimed to have ``improved naturalness'' compared with 
the pure SM.

These models are very interesting for several reasons, namely its relative 
simplicity and the fact that they can arise from certain SUSY scenarios 
(e.g. those with low-scale SUSY breaking \cite{Andrea,CEHI}), and likely 
from other kinds of models. Thus a closer look at them is appropriate. In 
particular, a more careful evaluation of the fine-tuning in these models 
and the subsequent comparison to the SM performance is needed. This is the 
first goal of this paper.

The second goal is to examine if the SM {\em alone} could be able to cope 
with the above-mentioned ``unexpected'' situations at LHC, i.e. a heavy SM 
Higgs and/or the possibility that LHC does not find any NP.

In section 2 we examine in detail the issue of naturalness of the pure SM 
(as low-energy effective theory). In particular we specify a sensible 
criterion to evaluate the degree of fine-tuning in the SM or in any other 
model (to allow a fair comparison). Also we discuss the role of radiative 
corrections, and compute the fine-tuning of the SM for different 
assumptions about $m_h$ and $\Lambda$. In section 3 we perform a similar 
analysis for the 2HDMs proposed in the literature with the aim of reducing 
the SM fine-tuning. We evaluate the fine-tuning and compare it with the SM 
one for each scenario. In section 4 we examine whether the SM alone could 
accommodate a heavy Higgs or a cut-off beyond the LHC reach with no 
fine-tuning price. Finally, in section 5 we present our conclusions. 
Besides, we collect in appendix~A the contributions to the $T$--parameter 
in generic 2HDMs, applying them to the three cases discussed in section~3.

\section{The Standard Model and the Scale of New Physics}

In order to evaluate the improvement in naturalness (if any) of generic 
2HDMs, or indeed of any alternative to the SM, it is necessary in 
the 
first place to perform an evaluation as rigorous as possible of the degree 
of fine-tuning of the SM itself for given values of the Higgs mass, $m_h$, 
and the cut-off scale, $\Lambda$.

First of all, eq.~(\ref{quadrdiv0}) can be renormalization-group (RG) 
improved, including leading-log corrections to all orders. In practice, 
this is taken care of by evaluating the couplings in eq.~(\ref{quadrdiv0}) 
at the cut-off 
scale, $\Lambda$ \cite{EJ}. Physically this makes good sense: one can 
think of $\delta_q m^2$ as a threshold correction from physics at 
$\Lambda$ and it naturally depends on couplings at that scale. Hence
\be
\label{BCform2}
m^2(\Lambda) = m_0^2 + \left.\delta_q m^2\right|_\Lambda\ ,
\ee
where $m_0^2$ is the tree-level value of the mass parameter
at the scale $\Lambda$ and
\be
\label{quadrdivL}
\left.\delta_q m^2\right|_\Lambda= {3\over 
64\pi^2}\Lambda^2\left[3g^2(\Lambda) 
+ g'^2 (\Lambda)
+ 8\lambda(\Lambda)
 - 8\lambda_t^2(\Lambda)\right]\ .
\ee

Second, for a fair comparison, one needs a sensible criterion to quantify 
the degree of fine-tuning, which should be applied to all the models.  
Here we follow the standard Barbieri and Giudice criterion \cite{BG}: we 
write the Higgs VEV as $v^2=v^2(p_1, p_2, \cdots)$, where $p_i$ are the 
fundamental input parameters of the model under study, and define 
$\Delta_{p_i}$, the fine tuning parameter associated to $p_i$, by
\be
\label{ftBG}
{\delta M_Z^2\over M_Z^2}= {\delta v^2\over v^2} = \Delta_{p_i}{\delta
p_i\over p_i}\ ,  \ee
where $\delta M_Z^2$ (or $\delta v^2$) is the change induced in
$M_Z^2$ (or $v^2$) by a change $\delta p_i$ in $p_i$. Roughly
speaking $1/|\Delta_{p_i}|$ measures the probability of a 
cancellation among terms of a
given size to obtain a result which is $|\Delta_{p_i}|$ times smaller. 
Due to the statistical meaning of $\Delta_{p_i}$, makes sense 
to define the total fine-tuning as \cite{CEHI}
\be
\label{Deltatot}
\Delta \equiv\left[\sum_i\Delta_{p_i}^2\right]^{1/2}\ .
\ee
(Other definitions, such as 
$\Delta = {\rm Max}\{\Delta_{p_i}\}$, are possible and have been
used in the literature.\footnote{Although in many cases both definitions 
give very similar results (typically one single $\Delta_{p_i}$ dominates 
$\Delta$) we believe definition (\ref{Deltatot}) is more satisfactory 
conceptually. As an (extreme) example consider the case of an observable 
$O$ that depends on a large number $N$ of input parameters, say 
$O=\sum_i\alpha_i p_i$ where $\alpha_i\sim {\cal O}(1)$ with 
random signs and with the measured value of $O$ and the natural 
values of the $p_i$ being of the same order. In such case all 
$\Delta_{p_i}^2\sim 1$ but the fine-tuning 
is ${\cal O}(\sqrt{N})$ (this example would correspond to a random walk 
where one expects such wandering of $O$ away from 1).})

In order to evaluate the fine-tuning $\Delta$ of the SM,
one should first identify the relevant unknown
parameters, $p_i$, to be plugged in eqs.~(\ref{ftBG}) and 
(\ref{Deltatot}). 
From eq.~(\ref{quadrdiv0}), we see that the most relevant ones are
$\Lambda$ and $\lambda$ (the experimentally well-known
couplings $g,g'$ and $\lambda_t$ have a negligible impact on the
fine-tuning \cite{CS}). Therefore, we will have
\be
\Delta\simeq \sqrt{\Delta_\Lambda^2+\Delta_\lambda^2}\ .
\ee 

\begin{figure}[t]
\vspace{1.cm} \centerline{
\psfig{figure=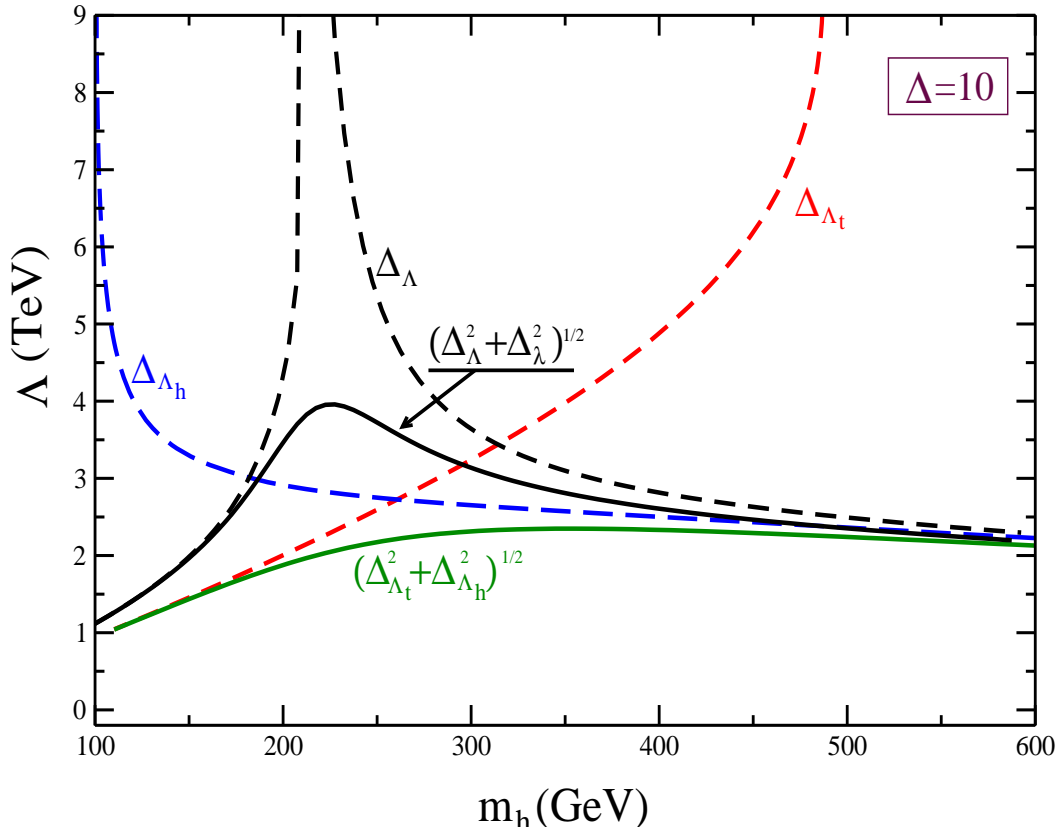,angle=-90,height=10cm,width=11cm,bbllx=4.cm,%
bblly=4.cm,bburx=20.cm,bbury=23.cm}
}
\caption{\footnotesize
SM naturalness upper bounds on the NP scale for $\Delta=10$ as a function 
of $m_h$  
with $\Delta=\Delta_{\Lambda_{t}}$ (red dashed), $\Delta_{\Lambda_{h}}$ 
(blue dashed), $(\Delta_{\Lambda_{t}}^2+\Delta_{\Lambda_{h}}^2)^{1/2}$
(solid green). The black lines show the corresponding bounds
for a single cut-off, $\Lambda$: $\Delta=\Delta_\Lambda$
(black dashed), $(\Delta_{\Lambda}^2+\Delta_{\lambda}^2)^{1/2}$
(black solid).
}
\label{fig:SM1}
\end{figure}

\begin{figure}[t]
\vspace{1.cm} \centerline{
\psfig{figure=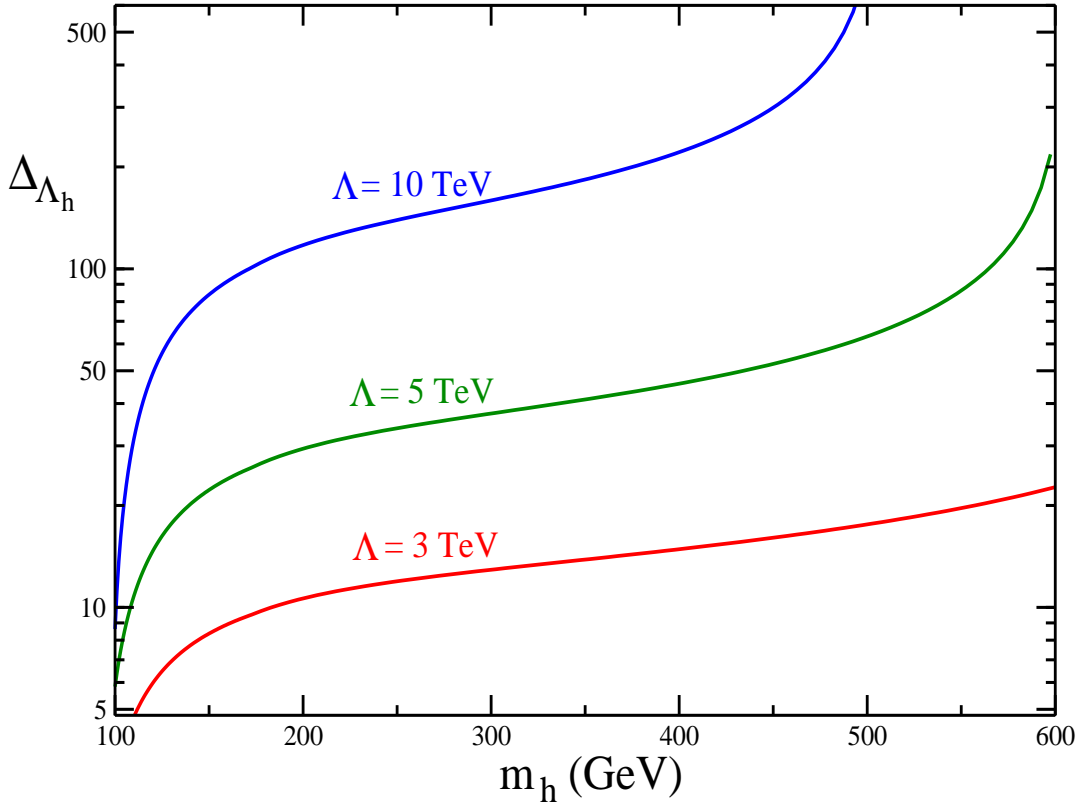,angle=-90,height=10cm,width=11cm,bbllx=4.cm,%
bblly=4.cm,bburx=20.cm,bbury=23.cm}
}
\caption{\footnotesize
SM fine-tuning associated to $\Lambda_h$ as a function of the Higgs mass 
for 
different values of the cut-off scale.}
\label{fig:SM1p}
\end{figure}

As stressed in \cite{BHR},
it could happen that the  NP that cancels the quadratically divergent 
corrections is different for the loops involving the top, the Higgs, etc. 
In that case, one should introduce different cut-offs:
\be
\label{quadrdif}
\delta_{\rm q} m^2 = {3\over 64\pi^2}\left[(3g^2 + g'^2)\Lambda_g^2
+ 8\lambda\Lambda_h^2 - 8\lambda_t^2\Lambda_t^2\right]\ .
\ee
In this case one should consider
$\Delta_{\Lambda_{t}}$ and $\Delta_{\Lambda_{h}}$ (the most relevant 
fine-tuning parameters) separately and use 
\be
\label{quadrature}
\Delta\simeq 
\sqrt{\Delta_{\Lambda_t}^2+\Delta_{\Lambda_h}^2+\Delta_\lambda^2}\ .
\ee
Fig.~\ref{fig:SM1} shows the naturalness upper bounds on $\Lambda$, 
derived for $\Delta_{\Lambda_{t}},  \Delta_{\Lambda_{h}}=10$, as a 
function of 
$m_h$ (red and blue dashed lines).
Notice that, keeping $\Lambda$ fixed, $\Delta_{\Lambda_{t}}$ decreases 
with 
increasing $m_h$ 
(or, equivalently, for fixed $\Delta_{\Lambda_{t}}$, the larger $m_h$, 
the larger 
may $\Lambda_t$ be). This follows trivially from
$v^2 = 
-m^2/\lambda$
and the one-loop expression for $\delta_q m^2$, eq.~(\ref{quadrdiv0}).
Then 
\be
\Delta_{\Lambda_{t}}\simeq {3\lambda_t^2\over 4\pi^2}
{\Lambda_t^2\over \lambda v^2}
= {3\lambda_t^2\over 2\pi^2}{\Lambda_t^2\over m_h^2}\ .
\ee
This fact has been used sometimes to suggest that a heavy Higgs behaves better
for naturalness than a light one. A similar reasoning would indicate that
$\Delta_{\Lambda_{h}}$ is independent of $m_h$: 
\be
\Delta_{\Lambda_h}\simeq - {3\lambda\over 4\pi^2}{\Lambda_h^2\over\lambda 
v^2}={3 \Lambda_h^2\over 4\pi^2v^2}\ .
\ee
RG effects cause important deviations from this 
tree-level expectation because the ratio of $\lambda$'s 
that enters $\Delta_{\Lambda_h}$ is roughly  
$\lambda(\Lambda)/\lambda(m_h)$. Using the RGE for $\lambda$
\be
\label{lambdaSM}
{d \lambda \over d \ln Q}={1\over 16\pi^2}\left[
24\lambda^2 +{3\over 
8}(3g^4+2g^2{g'}^2+{g'}^4)-6\lambda_t^4-3\lambda(3g^2+{g'}^2-4\lambda_t^2)
\right]\ ,
\ee 
it follows that, in the low $m_h$ range, 
$\lambda$ is small and runs to smaller values in the ultraviolet (UV), 
giving $\lambda(\Lambda)/\lambda(m_h)<1$ and weakening the naturalness 
bound on $\Lambda_h$, as shown in fig.~\ref{fig:SM1}. In the high $m_h$ 
range, $\lambda$ is bigger at $m_h$ and increases in the UV so that 
$\lambda(\Lambda)/\lambda(m_h)>1$, tightening the bound on $\Lambda_h$. 
The effect is stronger for larger $\Lambda$, see fig.~\ref{fig:SM1p}. The 
combined $\Delta= (\Delta_{\Lambda_{t}}^2+\Delta_{\Lambda_{h}}^2)^{1/2}$ 
interpolates between both behaviours: it is dominated by 
$\Delta_{\Lambda_t}$ at low $m_h$ and by $\Delta_{\Lambda_h}$ at large 
$m_h$ (see the solid green line $\Delta=10$ in Fig.~\ref{fig:SM1}). Notice 
that, although $\Lambda_{t}$ and $\Lambda_{h}$ are independent parameters, 
they are taken to be numerically equal in this figure.

On the other hand it is perfectly possible that
the NP that cancels the quadratic divergence associated 
to the top loop is the same that cancels that of the Higgs loops.
In other words, both could be different sectors of a single piece of NP, 
with $\Lambda_{t}\sim\Lambda_{h}$ (we discuss 
this in more detail in section~4). This can occur, for instance, when the 
NP 
corresponds 
to the supersymmetric partners of the SM particles, with all masses 
determined by a unique scale of SUSY breaking. Then one has to 
evaluate a single
$\Delta_\Lambda$; the corresponding contour plot $\Delta_\Lambda=10$
 is shown by the black dashed line in Fig.~\ref{fig:SM1}.
There we can notice a throat (sometimes called ``Veltman's throat'')
at $m_h\sim 225$ GeV, that arises from an accidental cancellation
between the various terms in eq.~(\ref{quadrdiv0}), in particular
between the top and Higgs contributions (which have opposite sign).
This fact has been used since long ago to suggest a preferred mass
for the Higgs \cite{veltman}. However, at present the Higgs mass 
(and thus the $\lambda$ coupling) is unknown and one has to include 
$\Delta_\lambda$ in the fine-tuning analysis \cite{CEHI}. The combined 
$\Delta=(\Delta_{\Lambda}^2+\Delta_{\lambda}^2)^{1/2}=10$ bound
is shown by the solid black line in Fig.~\ref{fig:SM1}, where one can 
see how the throat reduces to a bump. In summary, the solid lines of 
Fig.~\ref{fig:SM1} show the 
degree
of fine-tuning of the SM for given $\{\Lambda,m_h\}$ under
the assumption of independent\footnote{In the case of independent cut-offs
the total fine-tuning should also include the $\Delta_\lambda$ contribution,
but this does not modify substantially the result.} 
or correlated $\Lambda_{t}$ and
$\Lambda_{h}$ cut-offs.
Since the contributions to $\delta_q m^2$ proportional to $\Lambda_{t}^2$ 
and
$\Lambda_{h}^2$ have opposite signs, assuming a unique cut-off                  
 gives a less severe fine-tuning $\Delta\sim a\Lambda^2                    
- b \Lambda^2$ (with $a,b>0$) than for independent cut-offs,                    
$\Delta\sim \sqrt{(a\Lambda_{t}^2)^2                                          
+ (b \Lambda_{h}^2)^2}$, see eq.~(\ref{quadrature}).                            
(The opposite will be the case for contributions with the same sign,            
simply because $|a+b| > \sqrt{a^2+b^2}$.)   

\begin{figure}[t]
\vspace{1.cm} \centerline{
\psfig{figure=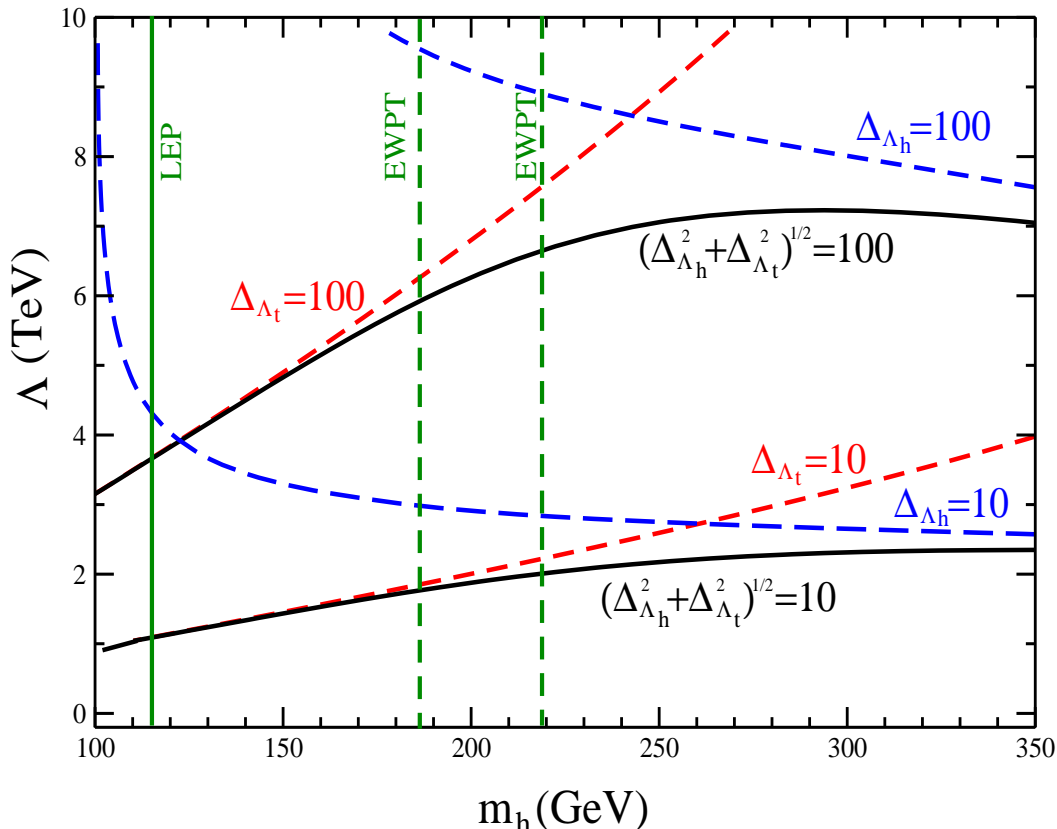,angle=-90,height=10cm,width=11cm,bbllx=4.cm,%
bblly=4.cm,bburx=20.cm,bbury=23.cm}
}
\caption{\footnotesize
SM naturalness upper bounds on the NP scale for $\Delta=10$ and 100 as a 
function of $m_h$ for independent $\Lambda_{t}$ and 
$\Lambda_{h}$ 
cut-offs; $\Delta=\Delta_{\Lambda_{t}}$ (red dashed), 
$\Delta_{\Lambda_{h}}$ 
(blue dashed), $(\Delta_{\Lambda_{t}}^2+\Delta_{\Lambda_{h}}^2)^{1/2}$
(solid black). The lower and upper limits on 
$m_h$ (from LEP and EWPT) are also shown.
}
\label{fig:SM2}
\end{figure}

\begin{figure}[t]
\vspace{1.cm} \centerline{
\psfig{figure=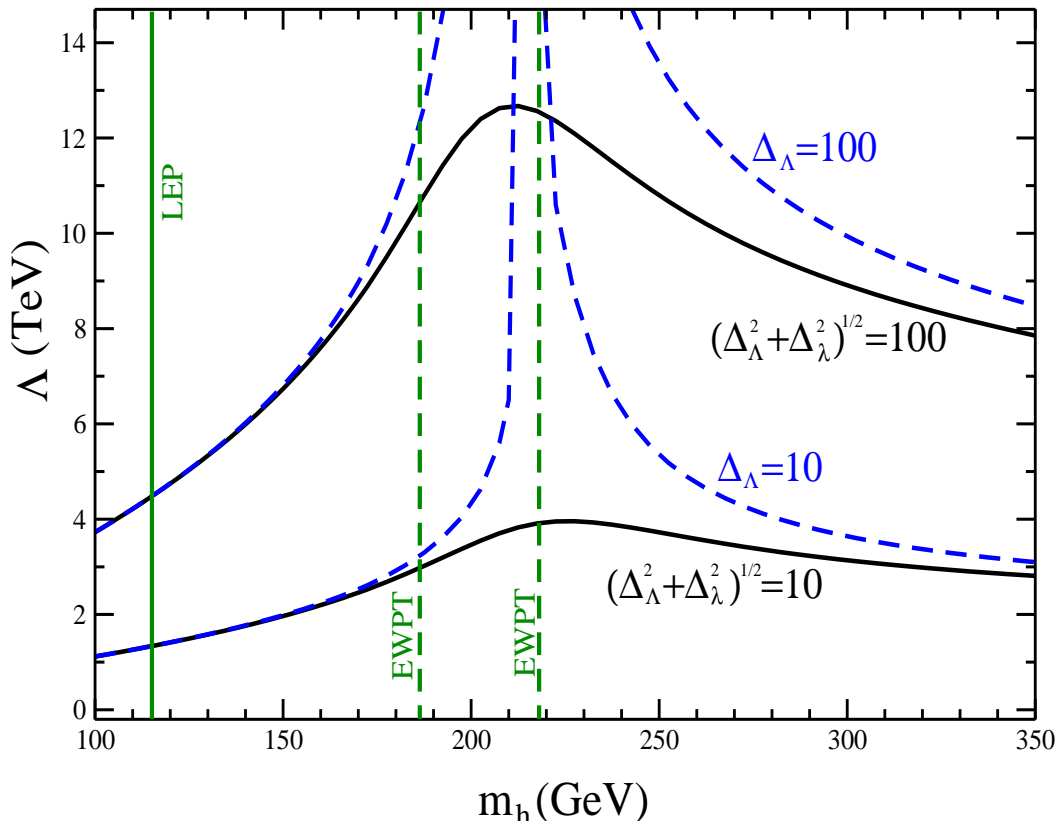,angle=-90,height=10cm,width=11cm,bbllx=4.cm,%
bblly=4.cm,bburx=20.cm,bbury=23.cm}
}
\caption{\footnotesize
Same as in Fig.~\ref{fig:SM2} but using a single cut-off $\Lambda$. 
$\Delta=\Delta_\Lambda$
(blue dashed lines), 
$(\Delta_{\Lambda}^2+\Delta_{\Lambda_{\lambda}}^2)^{1/2}$
(black solid lines).
}
\label{fig:SM3}
\end{figure}

It is useful, for future comparisons, to show the naturalness bounds for 
$\Delta=10$ and 100 in both cases (independent or correlated 
$\Lambda_{t}$ and
$\Lambda_{h}$). These are given in Fig.~\ref{fig:SM2} and 
Fig.~\ref{fig:SM3} respectively,
where we have also shown the present LEP lower limit on $m_h$ ($m_h\geq 
114.4$ GeV \cite{LEPmh}) and
the upper bounds from EWPT, $m_h<186$ GeV, or $m_h<219$ GeV if the LEP 
lower limit is included in the fit (both numbers at $95\%$ 
confidence-level) \cite{EWPTmh}. For later use we introduce as in 
\cite{BHR} the notation $m_{EW}$ for such EWPT upper bound on $m_h$.

As is clear from Figs.~\ref{fig:SM2} and \ref{fig:SM3}, the main 
difference between correlated and independent cut-offs is the presence or 
absence of the Veltman's throat (even if substantially reduced). Apart 
from this, both 
estimates of the fine-tuning show similar trends. Notice that choosing 
$\Lambda\sim 10$ TeV (to avoid problems with EWPT) requires $\Delta\simgt 
100$. If $\Lambda \sim 2$ TeV one generically has problems with EWPT, but 
this depends on the particular sort of NP showing up at that scale: there 
could be no problem in the presence of some symmetry (e.g. custodial 
symmetry) or a loop suppression (e.g. due to $R$-parity or $T$-parity, 
etc.). In such cases the Little Hierarchy problem would be absent or 
softened (i.e. there would be no tension between the values of $\Lambda$ 
required by EWPT and naturalness).

Let us finally discuss a technical point concerning our evaluation of
the various $\Delta_{p_i}$ in eq.~(\ref{ftBG}). From $v^2=-m^2/\lambda$
\be
{\delta v^2\over v^2}=
{\delta m^2\over m^2}-{\delta \lambda\over \lambda}\ .
\ee
Now, $\delta\lambda/\lambda$ is normally much smaller than $\delta m^2/ m^2$. 
In addition, only $\delta m^2/ m^2$ contains information about the 
quadratically divergent corrections,
which are the source of the hierarchy problem
on which we are focusing in this paper. 
So, to compute $\Delta_{p_i}$ in (\ref{ftBG}), we can approximate the 
tuning in $v^2$ by the tuning in $m^2$ using
\be
\label{vm}
{\delta v^2\over v^2}\simeq
{\delta m^2\over m^2}\simeq
\left.{\delta \{\delta_q m^2\}\over m^2}\right|_\Lambda\ ,
\ee
where we have evaluated ${\delta m^2/ m^2}$ at the  scale $\Lambda$
and used eq.~(\ref{BCform2}). This procedure is accurate, simplifies the 
computation and, furthermore, it makes sense since 
the actual cancellation between the tree-level and the radiative
contributions to $m^2$ occurs at the scale  $\Lambda$.

\section{Examples of Modified Higgs Sectors}

In this section we revisit three modifications of the Higgs sector that 
have been recently proposed in the literature with the general aim of 
improving the naturalness of the SM.

\subsection{The Inert Doublet Model (IDM)}

Motivated by the reduction of $\Delta_{\Lambda_t}$ for large $m_h$, 
Barbieri et al. have explored in \cite{BHR} the possibility of reconciling 
a large $m_h$ with EWPT in the most economical way. The 
model, 
called the ``Inert Doublet Model'', is a particular 2HDM with parity
\be
\label{duality}
H_2 \rightarrow -H_2\ .
\ee
This makes
$H_1$ the only Higgs field coupled to matter. The scalar potential is
\bea
\label{VM3}
V&=&\mu_1^2|H_1|^2\ +\ \mu_2^2|H_2|^2\ +\ \lambda_1|H_1|^4\ +
\ \lambda_2|H_2|^4\ +\ \lambda_3|H_1|^2|H_2|^2\ 
\nonumber\\
&+&\ \lambda_4|H_1^\dagger H_2|^2\ 
+\ {\lambda_5\over 2}[(H_1^\dagger H_2)^2\ +\ {\rm h.c.}] \ .
\eea
The next assumption is that the parameters of the potential are such that only
$H_1$ develops a VEV. $H_2$ does not couple to fermions and does not get 
a VEV
(thus the name ``inert doublet'') but it has weak and quartic 
interactions, playing an active role in EWPT. The Higgs VEV, $v=\sqrt{2}
\langle H_1^0\rangle $ is 
given by the SM-like relation
\bea
\label{VEV}
v^2 = -{\mu_1^2\over \lambda_1}\ .
\eea
We can write
\bea
\label{2H}
H_1 = \left(\begin{array}{c} G^+ \\
(v + h^0+i G^0)/\sqrt{2}
\end{array}     \right), \;\;\;
H_2 = \left(\begin{array}{c} H^+\\(H^0+iA^0)/\sqrt{2}
\end{array}      \right).
\eea
where $G^+, G^0$ are the usual SM Goldstones, $h^0$ is the SM-like Higgs 
and $H^+,H^0$ and $A^0$ are charged, scalar and pseudoscalar extra 
``Higgs'' states. The Higgs $h^0$ has a SM-like mass
\be
m_h^2= 2\lambda_1 v^2\ .
\ee
The states coming from the inert doublet $H_2$ have a common mass roughly 
given by $\mu_2^2$, with some splittings induced by electroweak symmetry 
breaking (EWSB):
\be
\label{massesM3}
m_I^2 = \mu_2^2 + \lambda_I v^2\ ,
\ee
with $I=\{H^+,H^0,A^0\}$ and
\bea
\label{massesM32}
\lambda_{H^+}&=&\lambda_3\ ,
\nonumber\\
\lambda_{H^0} &=&\lambda_3+\lambda_4+\lambda_5\ ,
\nonumber\\
\lambda_{A^0} &=& \lambda_3+\lambda_4-\lambda_5\ .
\eea
These mass splittings inside the $H_2$ multiplet break custodial 
symmetry and will play a crucial role in providing a contribution to 
$\Delta T$ 
of the right size to compensate that coming from a heavy $h^0$.
The SM Higgs gives a contribution to $T$ that grows logarithmically with 
$m_h$:
\bea
\label{T1}
T\sim - {3\over 8 \pi \cos^2 \theta_w}\ln{m_h\over M_Z}
\eea
which for $m_h\sim 400-600$ GeV is excluded at more than 99.9\% C.L. 
(The experimental value is $T^{\rm exp}\simeq 0.1\pm 0.15$).
The inert 
doublet provides an additional contribution (see appendix~A)
\bea
\label{T2}
\Delta T\simeq  {1\over 12 \pi s_W^2 M_W^2}(m_{H^+}-m_A)(m_{H^+}-m_S)\ ,
\eea
which, if $\Delta T \simeq 0.25\pm 0.1$, compensates the too negative 
contribution (\ref{T1}) getting agreement with experiment.
This requires \cite{BHR}
\bea
\label{T3}
(m_{H^+}-m_A)(m_{H^+}-m_S)=M^2,\;\;\; M=120^{+20}_{-30}\ {\rm GeV}\ ,
\eea
which seems to be a reasonable value for EWSB mass splittings. For later 
use, note that for $\mu^2_2\gg \lambda_I v^2$ the previous condition reads
\be
\label{DTrouble}
(\lambda_{H^+}-\lambda_{H^0})(\lambda_{A^0}-\lambda_{H^0}){v^4\over 
4\mu_2^2}=(\lambda_4^2-\lambda_5^2){v^4\over 4\mu_2^2}\simeq M^2\ .
\ee
This shows explicitly that the $\Delta T$ contribution of the inert 
doublet decouples with increasing $\mu_2^2$, as it should. More 
importantly for our purposes, it also shows that, for large $\mu_2^2$,
one needs large $\lambda_4^2-\lambda_5^2$ to get the right value of 
$\Delta T$.

This model is an interesting example of how to accommodate 
the possible detection of a heavy Higgs at the LHC. Moreover, the 
proponents of the model stress that having a heavy Higgs relaxes
the SM Hierarchy Problem, since it allows for larger $\Lambda_t$.
The price would be that the NP responsible for the cancellation
of the SM quadratic divergences could escape LHC (although we
could observe a modified Higgs sector). We would like to discuss these
naturalness aspects here. As shown in sect.~2,
a large $m_h$ does not necessarily imply less fine-tuning. In the SM 
context, although 
$\Delta_{\Lambda_t}$ decreases with $m_h$, $\Delta_{\Lambda_h}$ increases,
eventually dominating the total fine-tuning. This happens because
of the RG increase of $\lambda$ from $m_h$ to $\Lambda$,
where the quadratic radiative correction $\delta_q m^2$ has to be 
computed. In the present model something similar is likely to take 
place: let us focus for the moment on the impact of $\delta_q\mu_1^2$
on the fine-tuning.
From eq.~(\ref{VEV}), the relevant mass parameter for EW breaking
is $\mu_1^2$, which receives the quadratically-divergent radiative 
correction
\be
\label{quadrdiv3}
\delta_{\rm q} \mu_1^2 = {3\over 64\pi^2}\left[-8\lambda_t^2 \Lambda_t^2
+ (3g^2+g'^2)\Lambda_g^2 + 8\lambda_1\Lambda_{H_1}^2
+ {4\over 3} (2\lambda_3 +\lambda_4)\Lambda_{H_2}^2
\right]\ ,
\ee
where, following the assumption of ref.~\cite{BHR}, we have 
allowed 
independent cut-offs for different contributions. Eq.~(\ref{quadrdiv3})
has a structure which is very similar to the SM eq.~(\ref{quadrdivL}), 
with
the role of $\lambda$ played by $\lambda_1$. The two main differences are
the presence of the $\Lambda_{H_2}^2$ term and the RGE for $\lambda_1$, 
which is now:
\bea
{d \lambda_1 \over d \ln Q}&=&{1\over 16\pi^2}\left[
24\lambda_1^2 + \lambda_3^2 + (\lambda_3+\lambda_4)^2 + \lambda_4^2 
+\lambda_5^2\right.\nonumber\\
&&+\left.{3\over 8}(3g^4+2g^2{g'}^2+{g'}^4)-3\lambda_1(3g^2+{g'}^2)
\right]\ .
\eea 
The $\lambda_1$ and gauge terms in this RGE are as in the SM, but the 
additional terms cause $\lambda_1$ to grow with the scale more quickly 
than $\lambda$ in the SM [see eq.~(\ref{lambdaSM})].
Both differences contribute to produce a larger fine-tuning than in the SM 
for 
a given $m_h$ (although, of course, EWPT can now be under control for 
much higher $m_h$).
The best case, taking both effects into account, will 
occur for small $\lambda_{3,4,5}$. On 
the other hand, the $\lambda_i$ couplings cannot be chosen at will. As 
stressed above, they 
have to be consistent
with the desired $\Delta T$.
Besides this, there are additional constraints to ensure the stability of
the vacuum and the perturbativity of the $\lambda_i$ couplings \cite{BHR}. 

Following the analysis performed in \cite{BHR}, we choose the IDM 
parameters as follows. First we remind the reader that the lightest state 
in $H_2$ is stable (thanks to the $H_2\rightarrow -H_2$ symmetry) and is 
therefore a candidate for dark matter \cite{BHR}. This forces one to 
choose $m_{H^+}>m_{H^0},m_{A^0}$ so that this DM candidate is neutral. We 
follow \cite{BHR} and call $m_L={\mathrm 
min}\{m_{H^0},m_{A^0}\}$  the mass of this lightest state 
and $m_{NL}\equiv m_L+\Delta m={\mathrm max}\{m_{H^0},m_{A^0}\}$. Once we 
choose $m_L$ and $\Delta m$, $m_{H^+}$ is determined by eq.~(\ref{T3}). In 
fact, 
$\lambda_4$ and $\lambda_5$ are then fixed once the inert spectrum has 
been arranged in this way. On the other hand, $\lambda_1$ is determined by 
$m_h$. Of the two remaining quartic Higgs couplings we take 
$\lambda_2\simeq 0$ (its value is not really important) and  we choose 
$\lambda_3$ such that $\lambda_L\equiv 
\lambda_3+\lambda_4-|\lambda_5|$ satisfies perturbativity and vacuum 
stability constraints \cite{BHR}. For numerical work we then take $M=120$ 
GeV, $\Delta m=50$ GeV and $\lambda_L=-0.5$. This choice gives us a wide 
range for $m_L$ to agree with the needed $\Delta T$ and to meet all other 
constraints \cite{BHR}.

\begin{figure}[t]
\vspace{1.cm} \centerline{
\psfig{figure=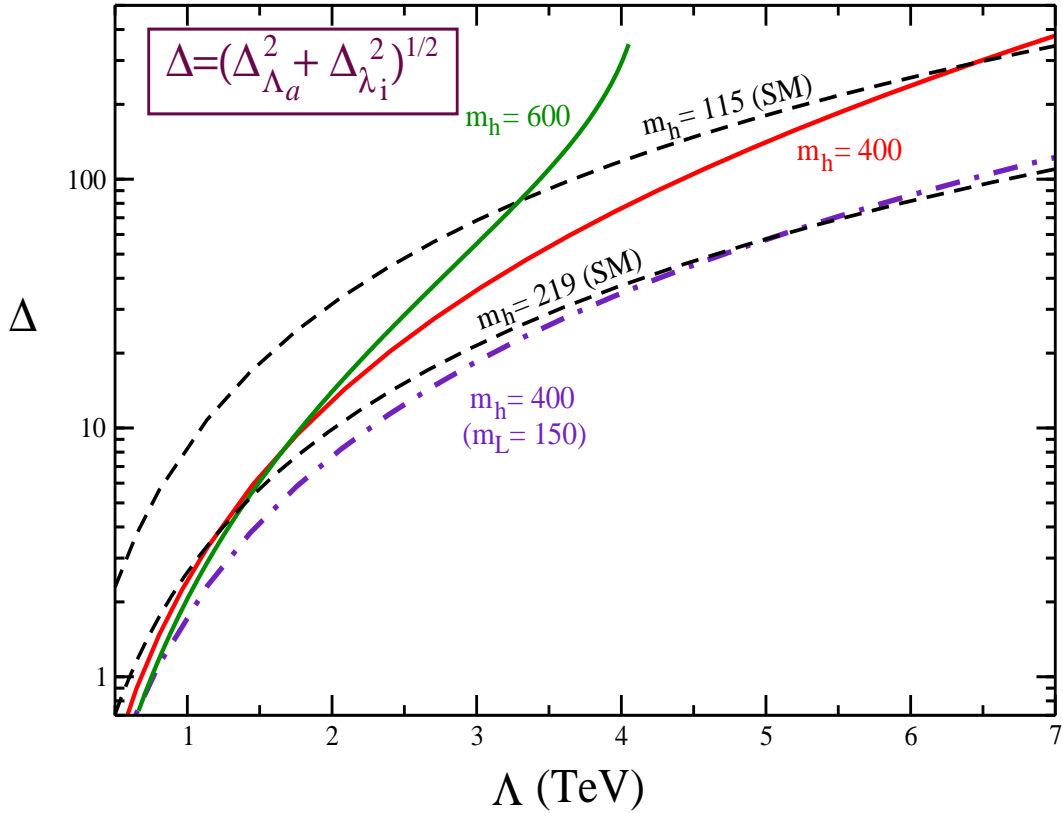,angle=-90,height=10cm,width=11cm,bbllx=4.cm,%
bblly=4.cm,bburx=20.cm,bbury=23.cm}
}
\caption{\footnotesize
Total fine-tuning $\Delta$ versus the cut-off $\Lambda$
in the Inert Doublet Model for 
$m_h= 400$
and 600 GeV using $m_L = m_h$ (red and green solid lines); and $m_h=400$ 
GeV, $m_L = 150$ GeV (violet dashed-dotted). The dashed black lines show 
the
total SM fine-tuning for $m_h=115$ GeV and 219 GeV, using 
independent cut-offs.
}
\label{fig:Inert1}
\end{figure}

\begin{figure}[t]
\vspace{1.cm} \centerline{
\psfig{figure=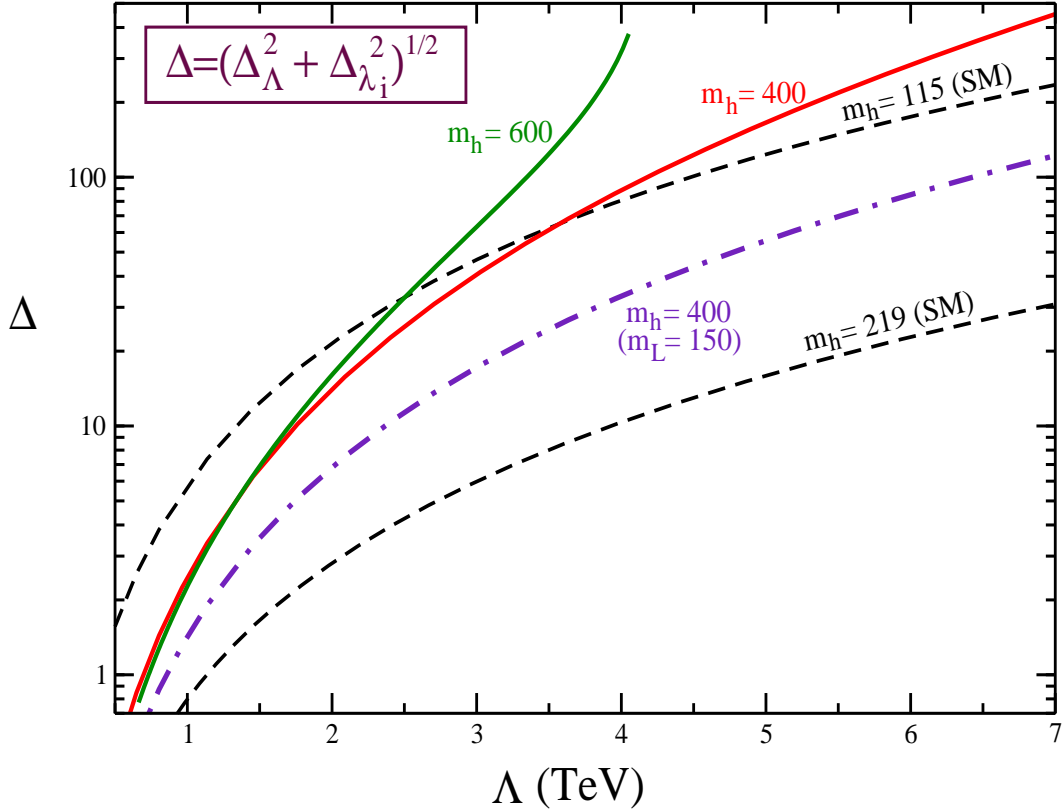,angle=-90,height=10cm,width=11cm,bbllx=4.cm,%
bblly=4.cm,bburx=20.cm,bbury=23.cm}
}
\caption{\footnotesize
Same as in Fig.~\ref{fig:Inert1}, but now using a unique cut-off, 
$\Lambda$.
}
\label{fig:Inert2}
\end{figure}

Fig.~\ref{fig:Inert1} shows the total fine-tuning\footnote{We evaluate 
$\Delta$ by approximating the tuning in $v^2$ by the tuning in $\mu_1^2$ 
in the spirit of eq.~(\ref{vm}) for the SM.} $\Delta = (\sum_{a=t,H_1,H_2} 
\Delta_{\Lambda_a}^2 
+\sum_i\Delta_{\lambda_i}^2)^{1/2}$ versus the cut-off $\Lambda$ (for 
simplicity we show the results when all the cut-offs are numerically 
equal) for $m_h= 400$ and 600 GeV (red and green solid lines) for the 
particular case $m_L = m_h$. Notice that the case with $m_h=400$ GeV 
behaves better, as expected from the general trend discussed in 
section~2 (see Fig.~\ref{fig:SM1p}). The violet 
dashed-dotted line corresponds to $m_h=400$ GeV and $m_L = 150$ GeV. This 
is 
a case with small $\lambda_{3,4,5}$ leading to a fine-tuning that is 
sensibly smaller, in agreement with the above discussion. In fact, this 
line is close to the optimal situation in this scenario, see below. In any 
case, it is clear that reaching $\Lambda>2$ TeV requires a substantial 
fine-tuning ($\Delta>10$). To see if this model improves the situation 
over the SM one, we have also plotted the SM fine-tuning $\Delta = 
(\Delta_{\Lambda_t}^2+ \Delta_{\Lambda_h}^2+\Delta_{\lambda}^2)^{1/2}$ 
as a band between  
the limit cases $m_h=115$ GeV and 219 GeV (dashed black lines). Clearly, 
the 
situation of the IDM can be hardly considered an improvement in 
naturalness, especially if the SM Higgs is near its upper EWPT bound. This 
conclusion is strengthened if one assumes a universal cut-off for all the 
contributions in $\delta_q m^2$ and $\delta_q \mu_1^2$ (for the SM and 
IDM respectively). As discussed in 
sect.~2 this is a perfectly reasonable possibility. The corresponding 
fine-tuning curves are shown in Fig.~\ref{fig:Inert2}, where the SM (for 
$m_h$ around 200 GeV) is in a much better position, even when the optimal 
choice $m_L = 150$ GeV is used for the IDM.

Our conclusion at this point is therefore that, although the IDM could 
offer an appealing explanation for the hypothetical detection of a heavy 
Higgs, it does not improve the naturalness of EWSB with respect to the SM. 
This conclusion would be softened if i) one assumes independent cut-offs, 
or ii) if one requires $\Delta$ to be ${\cal O}(1)$. Notice that in the 
latter case $\Lambda$ is smaller, and then the effect of the RG running on 
$\Delta_{\Lambda_H}$ (which is harmful for the fine-tuning) is less 
important. Both assumptions were taken in ref.~\cite{BHR}. However, we 
believe that to require $\Delta=1$ is excessively severe, especially 
taking 
into account that a small $\Lambda$ generically leads to problems with 
EWPT.

\begin{figure}[t]
\vspace{1.cm} \centerline{
\psfig{figure=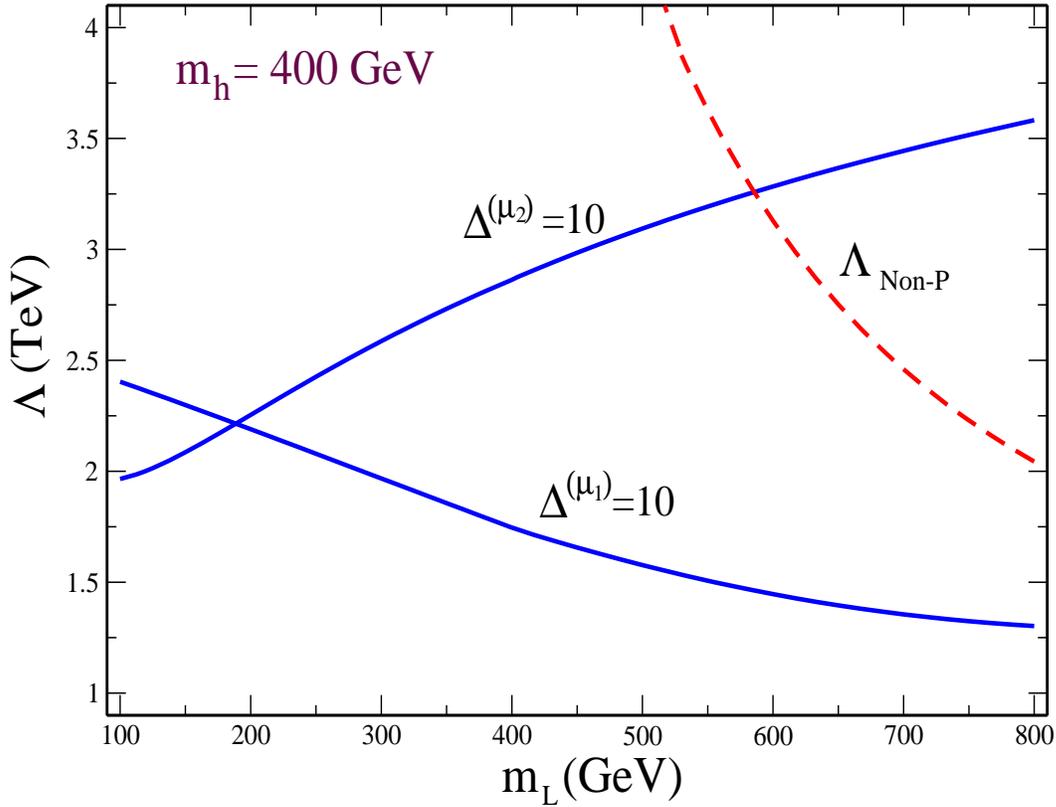,angle=-90,height=10cm,width=11cm,bbllx=4.cm,%
bblly=4.cm,bburx=20.cm,bbury=23.cm}
}
\caption{\footnotesize
IDM naturalness upper bounds on $\Lambda$ for $\Delta=10$ as a function 
of $m_L$ with $\Delta=\Delta^{(\mu_{1})},\ \Delta^{(\mu_{2})}$  
(solid blue). The red dashed line shows the limit of perturbativity 
($\lambda_1= 4\pi$).
}
\label{fig:Inert3}
\end{figure}

Let us now examine how the fine-tuning in $\mu_1^2$ changes 
with $m_L$: 
this is shown in Fig.~\ref{fig:Inert3} for $m_h=400$ GeV. Naturalness 
clearly favours low values of $m_L$. This is because high $m_L$ means high 
$\mu_2$ and, in order to satisfy the $\Delta T$ constraint, larger values 
of $\lambda_4$ and/or $\lambda_5$ are needed [see eq.~(\ref{DTrouble})]. 
This then enhances $\delta_q\mu_1^2$ [see eq.~(\ref{quadrdiv3})] and thus 
$\Delta^{(\mu_1)}$. We can 
also see from this plot that choosing a large $m_h$ (and thus a large 
$\lambda_1$) limits the scale up to which we can maintain perturbativity. 
The 
dashed line shows this limit (for $\lambda_1=4\pi$).

The $\delta_q \mu_1^2$ contribution
is not the only source of potential fine-tuning in the IDM:
$\delta_q \mu_2^2$, which is given by
\be
\label{quadrdiv4}
\delta_{\rm q} \mu_2^2 = {3\over 64\pi^2}\left[
 (3g^2+g'^2)\Lambda_g^2 + 8\lambda_2\Lambda_{H_2}^2
+ {4\over 3} (2\lambda_3 +\lambda_4)\Lambda_{H_1}^2
\right]\ ,
\ee
can also be an additional source of (independent)
fine-tuning. Let us call the corresponding fine-tunings
$\Delta^{(\mu_1)}$ and $\Delta^{(\mu_2)}$. 
Notice from (\ref{massesM3}) that a small $m_L$ 
[the optimal choice for $\Delta^{(\mu_1)}$] 
typically requires small $\mu_2^2$. Then, the larger
$\delta_q \mu_2^2$, the more unnatural this possibility is.
Fig.~\ref{fig:Inert3} shows $\Delta^{(\mu_2)}$, evaluated in 
the same way as $\Delta^{(\mu_1)}$. As expected, for $\Lambda$ fixed, 
$\Delta^{(\mu_2)}$ increases
for decreasing $m_L$, which balances the behavior of $\Delta^{(\mu_1)}$.
In fact one should multiply both fine-tunings, as they correspond
to different quantities, but, even if one does not, it is clear that the
improvement gained in $\Delta^{(\mu_1)}$  by going to small $m_L$
is lost by this additional source of fine-tuning in $\mu_2$.

From the previous discussion we finally conclude that i) although the IDM 
model is very interesting, it does not improve the naturalness of the 
SM; ii) the structure of the model requires an additional fine-tuning in 
$\mu_2^2$ of a size that can be similar to that required for a correct EW 
breaking (i.e. the tuning associated to $\mu_1^2$).
This result is common in models with a structure more 
complicated than that of the SM. Normally, epicycles are penalized in 
naturalness estimates (which is somehow satisfactory). 
This happens e.g. in Little Higgs Models
\cite{CEHII} and also here, though in a much less dramatic way.
Therefore the possibility that the NP responsible for the
cancellation of the dangerous quadratic divergences could escape LHC 
detection is similar in the IDM and in the SM. In the IDM case,
however, a Higgs sector different from the SM one would be 
observed.

\subsection{The Barbieri-Hall Model}

This model (BH Model from now on), presented in ref.~\cite{BH},
is another particular version of a 2HDM aimed at improving the 
naturalness of the SM. The idea here is to maintain the
lightest Higgs below the EWPT bound
($m_h \simlt 219$ GeV), but coupling
the top quark mainly to the heaviest Higgs, so that 
$\Delta_{\Lambda_t}$ can be much smaller than in the SM.

The Higgs potential has the form (\ref{VM3}), but now the 
parameters are chosen so that both $H_1$ and $H_2$
get VEVs: $\langle H^0_i \rangle = v_i/\sqrt{2}$, with $v_1^2 + v_2^2 = 
(246\ {\rm GeV})^2$. More precisely, the minimization conditions
read
\bea
\label{mincond}
\mu_1^2+\lambda_1 v_1^2 + {1 \over 2}\tilde\lambda v_2^2&=&0\ 
,
\nonumber\\
\mu_2^2+\lambda_2 v_2^2 + {1 \over 2}\tilde\lambda v_1^2&=&0\ ,
\eea
where $\tilde\lambda\equiv\lambda_3 +\lambda_4 +\lambda_5$.
In addition, a discrete symmetry is imposed
so that only $H_2$ couples to the up-quarks.
The squared mass matrix for the two neutral
Higgs bosons is
\be
\label{massmatrix}
\left(\begin{array}{cc} 
2\lambda_1 v_1^2&  \tilde\lambda v_1v_2\\
 \tilde\lambda v_1v_2&2\lambda_2 v_2^2
\end{array}\right)\ .
\ee
Assuming that
the 22 entry is the largest and the off-diagonal entry is small, the two 
mass eigenvalues $m_\pm^2$ are
\be
\label{masseigen}
m_+^2\simeq 2\lambda_2v_2^2,\;\;\;\; m_-^2 \simeq  
2\left(\lambda_1 - {{\tilde\lambda}^2\over 4\lambda_2}\right)v_1^2\ .
\ee
The important point is that, in this case, the lightest and the heaviest
neutral Higgs bosons are mainly along the $h_1$ and $h_2$ directions
respectively: $h_-= \cos\alpha\ h_1 +\sin\alpha\ h_2 $,
$h_+= \cos\alpha\ h_2 -\sin\alpha\ h_1 $ with a small mixing angle:
\be
\alpha \simeq {-\tilde\lambda\over 2\lambda_2\tan \beta}\ ,
\ee
where $\tan \beta = v_2/v_1$. Therefore, the lightest Higgs, $h_-$,
has almost no coupling to the up-quarks, and in particular to the top.
The quadratically divergent corrections to the $\mu_i^2$ mass
parameters are
\bea
\label{quadrdivBH}
\delta_{\rm q} \mu_1^2 &=& {3\Lambda^2\over 64\pi^2}\left[
(3g^2+g'^2) + 8\lambda_1+ {4\over 3} (2\lambda_3 +\lambda_4)
\right]\ ,
\nonumber\\
\delta_{\rm q} \mu_2^2 &=& {3\Lambda^2\over 64\pi^2}\left[
(3g^2+g'^2) + 8\lambda_2+ {4\over 3} (2\lambda_3 +\lambda_4)
-8\lambda_t^2
\right]\ ,
\eea
where we have taken a universal cut-off for simplicity, but
each term can be multiplied by a different cut-off if desired.
It is explicit from (\ref{quadrdivBH}) that $\lambda_t$ only affects 
$\mu_2^2$. 

Let us  focus first on the impact of $\delta_{\rm q} \mu_2^2$ on the 
fine-tuning, as the authors of \cite{BH} do. From
eqs.~(\ref{mincond}) and the smallness of 
$\alpha$,  we see that $v_2^2=v^2\sin^2\beta\simeq -\mu_2^2/\lambda_2$.
So, the fine-tuning associated to $\Lambda_t$ is given by
\be
\label{Deltat}
\Delta_{\Lambda_{t}}
\simeq\left|{\Lambda_t\over \mu_2^2} 
{\delta(\delta_q\mu_2^2)Â\over\delta\Lambda_t}
\right|_\Lambda
\simeq {3\lambda_t^2\over 4\pi^2}
{\Lambda_t^2\over \lambda_2 v_2^2}
\simeq {3\lambda_t^2\over 2\pi^2}{\Lambda_t^2\over m_+^2}\ .
\ee
Hence, $\Delta_{\Lambda_{t}}$ is suppressed for large
$m_+$, even if $h_-$ is light.
This trick was used in ref.~\cite{BH} to push $\Lambda_t$
to quite high values: taking $m_+ \simeq 500-1000$ GeV,
$\Lambda_t$ can be as large as 2 TeV, even if one
demands $\Delta_{\Lambda_{t}}\leq 1$.

The previous analysis, however, does not take into account the 
contribution $\Delta_{\Lambda_{H_2}}$ to the total fine-tuning. 
This effect was not considered in \cite{BH} but it is the one that puts 
the strongest constraint on the 
scale of NP.
Due to the 
large size assumed for
$m_+$, $\lambda_2$ is large and gets even larger in the UV, through RG 
running. As we have discussed above, the RG 
increase of $\lambda_2$ from $m_+$ to $\Lambda$ enhances the corresponding 
contribution to $\delta_q \mu_2^2$, and thus the value of 
$\Delta_{\Lambda_{H_2}}$.
 Actually, beyond some scale $\Lambda_{\rm Non-P}$ not too far 
from the EW scale, $\lambda_2$ gets non-perturbative (say $\lambda_2\geq 
4\pi$). Beyond $\Lambda_{\rm Non-P}$ the model enters a strong-coupling 
regime and eventually $\lambda_2$ reaches a Landau Pole at some higher 
scale. The perturbative limit on $\Lambda$ is, for some choices of the 
parameters, even below the previous estimates of $\Lambda_t$. In any case, 
we are interested in the cut-off scale that produces a total fine-tuning 
$\Delta\leq 10$ (or any other sensible value) and this typically will 
happen below $\Lambda_{\rm Non-P}$.

Of course, the effect of the RGE for $\lambda_2$ depends on 
the values of other couplings, especially on $\lambda_{3,4,5}$.
Explicitly,
\bea
\label{RGEl2}
{d \lambda_2 \over d \ln Q}&=&{1\over 16\pi^2}\left[
24\lambda_2^2 + \lambda_3^2 + (\lambda_3+\lambda_4)^2 +\lambda_5^2\right.
\nonumber\\
&&+\left.{3\over 
8}(3g^4+2g^2{g'}^2+{g'}^4)-6\lambda_t^4-3\lambda_2(3g^2+{g'}^2-4\lambda_t^2)
\right]\ .
\eea
Notice that the extra couplings always contribute to strengthen
the RG increase of $\lambda_2$, so a good choice for fine-tuning
purposes is to minimize their effect by taking their values as
small as possible. This choice also minimizes the 
quadratically-divergent
corrections, $\delta_{\rm q} \mu_{1,2}^2$, as given by 
eqs.~(\ref{quadrdivBH}).
Taking into account that the masses of the charged
and the pseudoscalar Higgs are given by
\bea
\label{massesHA}
m_{H^+}^2 &=& -{1\over 2}(\lambda_4 + \lambda_5) v^2\ ,
\nonumber\\
m_{A^0}^2 &=& -\lambda_5 v^2\ ,
\eea
it seems that $\lambda_3=\lambda_4=0$ might be an optimal choice.
Then $\lambda_5$ must be negative, with a lower bound (in absolute size)
given by the lower bound on $m_{H^+}$. For $\tan \beta= 0.8-1$ \cite{BH} 
this bound, coming from $b\rightarrow s\gamma$ constraints, is about $200-250$ 
GeV \cite{bsg}. 

\begin{figure}[t]
\vspace{1.cm} \centerline{
\psfig{figure=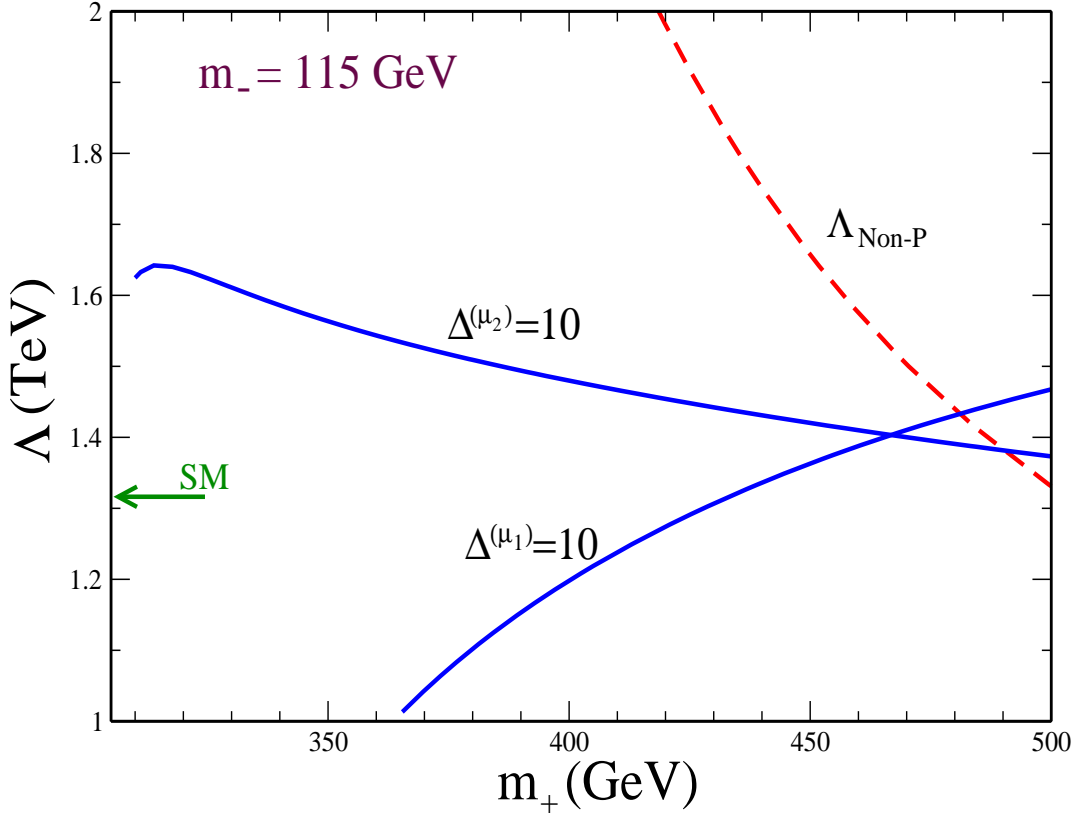,angle=-90,height=10cm,width=11cm,bbllx=4.cm,%
bblly=4.cm,bburx=20.cm,bbury=23.cm}
}
\caption{\footnotesize 
Naturalness upper bounds for $\Delta^{(\mu_i)}=10$ as a function of $m_+$ 
in the BH Model 
with $\lambda_3=\lambda_4=0$   
(solid blue). The $\Lambda_{\rm Non-P}$ (red dashed) line
gives the perturbativity limit ($\lambda_2= 4\pi$). The SM arrow 
corresponds to $m_h=115$ GeV.}
\label{fig:BH1}
\end{figure}

Using a unique cut-off, $\Lambda_{t}\equiv \Lambda_{H_i}\equiv \Lambda$, 
the naturalness upper bound on $\Lambda$, as a function of $m_+$ for 
$m_-=115$ GeV is shown in Fig.~\ref{fig:BH1}, for 
$\Delta^{(\mu_2)}=\{[\Delta^{(\mu_2)}_\Lambda]^2 
+\sum_i[\Delta^{(\mu_2)}_{\lambda_i}]^2\}^{1/2}=10$. The perturbativity 
limit, $\Lambda_{\rm Non-P}$, is also shown. The origin of the lower bound 
$m_+\simgt 310$ GeV in Fig.~\ref{fig:BH1} is the following. From the mass 
matrix (\ref{massmatrix}) it follows that for a given value of $m_-$, 
there is a minimal value of $m_+$, which occurs for $\lambda_2 
v_2^2=\lambda_1 v_1^2$ and $|\tilde\lambda|$ as small as possible. More 
precisely
\be 
(m_+^2)_{\rm min}=m_-^2+2|\tilde\lambda|v_1 v_2\ . 
\ee 
In our case, with $\tan\beta=1$ and $\lambda_3=\lambda_4=0$ we get 
\be 
(m_+^2)_{\rm min}=m_-^2+2 m_{H^+}^2\ . 
\ee 
which, for $m_-=115$ GeV and $m_{H^+}=200$ GeV gives $m_+\simgt 310$ GeV.
We have checked that with our choice of parameters $T$ is in agreement 
with the experimental value in all the range shown for $m_+$. In fact, for 
the largest values of $m_+$ a cancellation similar to that in the IDM 
(between $\Delta T_\parallel$ and $\Delta T_\perp$, see appendix~A) is at 
work.

For comparative purposes we also show in Fig.~\ref{fig:BH1} the 
naturalness bound in the SM for the same Higgs mass (see 
Fig.~\ref{fig:SM2}). Clearly, there is no substantial improvement with 
respect to the SM. Furthermore, one should take into account the 
fine-tuning associated to $\mu_1^2$, {\it i.e.} $\Delta^{(\mu_1)}$. 
Actually, it is clear from the same figure that, although 
$\Delta^{(\mu_2)}$ decreases with $m_+$ (for fixed $\Lambda$), 
$\Delta^{(\mu_1)}$ goes the opposite way. In fact, $\Delta^{(\mu_1)}$ is 
very restrictive, even more than $\Delta$ in the SM for $m_h=m_-$. The 
reasons for this behaviour 
are, first that having $\tilde\lambda\neq 0$ forces $\lambda_1(m_h)$ to be 
larger than $\lambda$ for the same $m_-=115$ GeV [see 
eq.~(\ref{masseigen})]. 
Second, the RG evolution 
of $\lambda_1$ is very different from that of $\lambda$ precisely because 
$H_1$ does not couple to the top quark: while in the SM $\lambda$ gets 
much weaker in the UV, $\lambda_1$ does not change much. These effects 
cause the $\lambda_1$ contribution in $\delta_q\mu_1^2$ to be much larger 
than in the SM for the same $m_-=115$ GeV.
The global situation 
is then similar to that in the IDM (see Fig.~\ref{fig:Inert3}): the 
improvement 
gained in $\Delta^{(\mu_2)}$ by going to small $m_+$ is counterbalanced by 
the $\Delta^{(\mu_1)}$ fine-tuning. Again, one should multiply both 
fine-tunings, but even if we do not (i.e. if we are conservative) it is 
clear how models with more structure are penalized in naturalness 
considerations.

\begin{figure}[t]
\vspace{1.cm} \centerline{
\psfig{figure=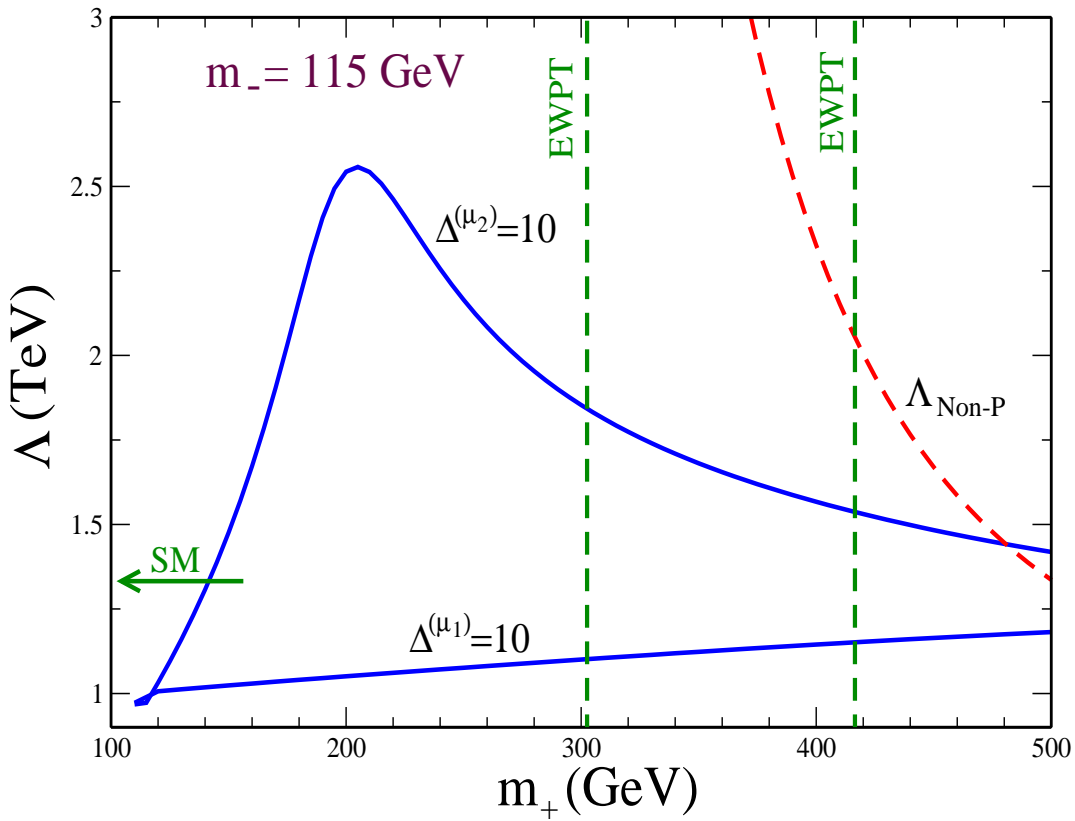,angle=-90,height=10cm,width=11cm,bbllx=4.cm,%
bblly=4.cm,bburx=20.cm,bbury=23.cm}
}
\caption{\footnotesize
Same as in Fig.~\ref{fig:BH1}, but using $\lambda_4=\lambda_5$ and 
$\tilde\lambda=0$.
The two green dashed lines correspond to the $m_+$ upper bounds
from EWPT.
}
\label{fig:BH2}
\end{figure}

One can try to change the $\lambda_3=\lambda_4=0$ assumption 
in order to relax the lower bound on $m_+$ and to reduce the fine-tuning.
A convenient choice is to 
use $\lambda_4=\lambda_5$ and adjust $\lambda_3$ to get $\tilde\lambda=0$.
This allows to minimize the splitting between $m^2_-$ and $m^2_+$. The 
result for the tuning is shown in Fig.~\ref{fig:BH2}. By going to lower 
values of $m_+$ we access a region of parameter space where a 
Veltman-like cancellation takes place in $\Delta^{(\mu_2)}$ between the 
quadratically divergent contributions of the top and the Higgses. On the 
other hand, the 
tuning $\Delta^{(\mu_1)}$ is very similar to the SM one but slightly 
worse: now, in addition to the different RG-evolution of $\lambda_1$ 
explained above, $\lambda_3$ is not small and also contributes to 
$\delta_q\mu_1^2$. (In fact, one cannot make $\tilde\lambda=0$ with 
$\lambda_3=0$ due to the lower bound on§ $m_{H^+}$.) Once more it 
is clear that the improvement 
gained in $\Delta^{(\mu_2)}$ by going to small $m_+$ is counterbalanced by 
the $\Delta^{(\mu_1)}$ fine-tuning.

With $\lambda_4=\lambda_5$ one has $m_A=m_{H^+}$ and the Higgs 
contribution to $T$ is very simple (see appendix~A). The constraint from 
EWPT can then be written as
\cite{BH} 
\be 
m_+^2<m_{EW}^2\left[{m_{EW}^2\over m_-^2}\right]^{1/\tan^2\beta}\ . 
\ee 
For $m_{EW}=\{186,219\}$ GeV and $m_-=115$ GeV, this gives 
$m_+<\{301,417\}$ GeV and these bounds are
represented as well in the figure. 

\begin{figure}[t]
\vspace{1.cm} \centerline{
\psfig{figure=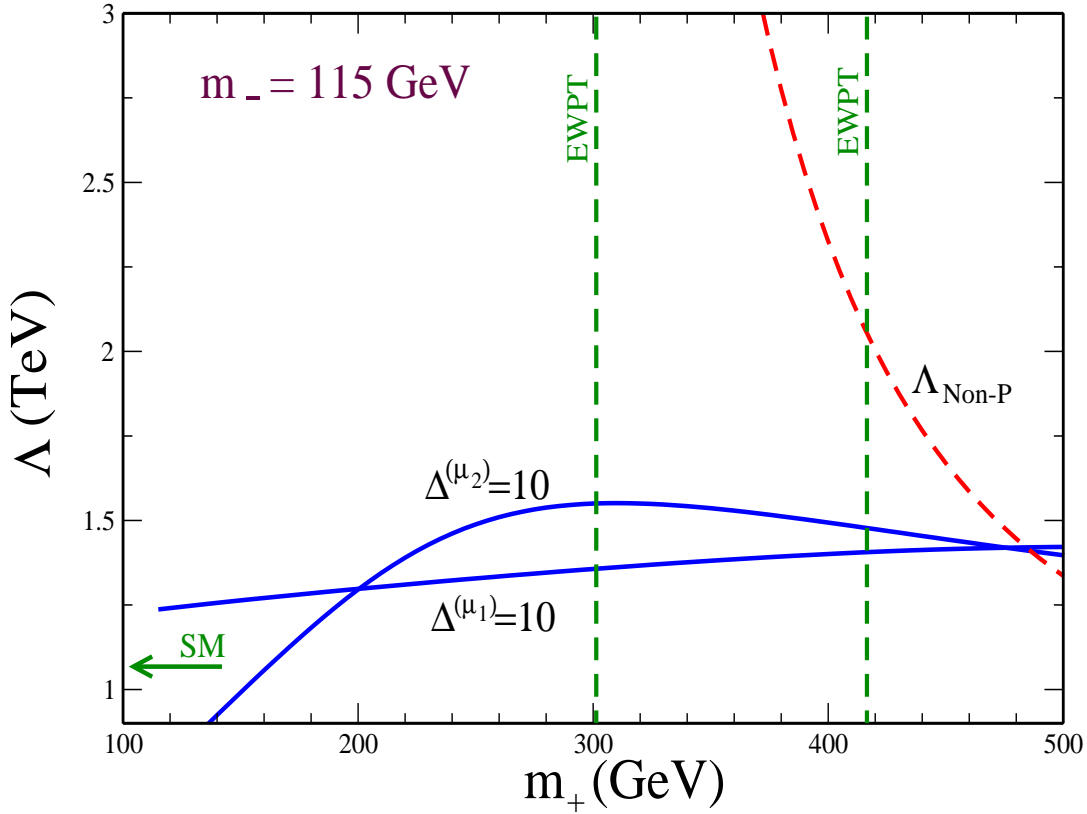,angle=-90,height=10cm,width=11cm,bbllx=4.cm,%
bblly=4.cm,bburx=20.cm,bbury=23.cm}
}
\caption{\footnotesize Same as in Fig.~\ref{fig:BH2} but with independent 
cut-offs $\Lambda_t$ and $\Lambda_{H_i}$.
}
\label{fig:BH3}
\end{figure}

Finally, one can calculate the fine-tuning using uncorrelated 
$\Lambda_{t}$ and $\Lambda_{H_i}$ cut-offs. This does not improve the 
naturalness since the accidental cancellation in $\Delta^{(\mu_2)}$ around 
$m_+ \sim 200$ GeV is now absent. In fact, according to the general 
discussion in section~1 we expect that the bound from $\Delta^{(\mu_2)}$ 
($\Delta^{(\mu_1)}$) will be stronger (weaker) because $\delta_q\mu_2^2$ 
($\delta_q\mu_1^2$) contains contributions of opposite (the same) sign. 
This is illustrated in Fig.~\ref{fig:BH3}, where the parameters of the 
model have been taken as for Fig.~\ref{fig:BH2}. In Fig.~\ref{fig:BH3} we 
have used the same numerical values for $\Lambda_{t}$ and $\Lambda_{H_i}$, 
but one could take different values for the different cut-offs. Then, 
$\Lambda_t$ could be much higher than usual, as already discussed around 
eq.~(\ref{Deltat}). So, the NP responsible for cancelling the top 
quadratic divergences, presumably strongly interacting, could be beyond 
LHC reach, as stressed in ref.~\cite{BH}. However, the NP that compensates 
the large quadratic corrections associated to the Higgs sector itself 
should show up at much lower scales\footnote{In that circumstance, one 
should worry about how rigorous is it to use the SM as the effective 
theory between $\Lambda_H$ and $\Lambda_t$. These results at least 
indicate that the strongly coupled NP might appear at higher scales than 
other kinds of NP.}.

In summary, this 2HDM shows that $\Lambda_t$ could be much larger
than $\Lambda_H$ even if the light Higgs is within the experimentally
preferred range. However, the global fine-tuning is not improved and
we generically expect NP (beyond the 2HD sector) to be on the LHC reach.

\subsection{Twin Higgs Model}

This model, originally proposed by Chacko, Goh and Harnik in 
ref.~\cite{TH}, postulates the existence of a mirror world: a $Z_2$ 
replica of the full SM. Calling $H_1$ the SM Higgs and $H_2$ its mirror 
copy, 
the Higgs sector of this model is a very particular kind of 2HDM with 
potential
\be
V=\mu^2(|H_1|^2+|H_2|^2)+\lambda(|H_1|^2+|H_2|^2)^2+\gamma 
(|H_1|^4+|H_2|^4)\ ,
\label{VTH}
\ee
that respects the $Z_2$ parity but allows communication with the mirror 
world \cite{wil} through a mixed term $|H_1|^2|H_2|^2$. A discussion of 
the naturalness of EWSB in this model was performed by ref.~\cite{TH} and 
later on by \cite{BGH}. For $\gamma>0$ 
and $\mu^2<0$ this potential has a minimum that breaks the electroweak 
symmetry with $\langle H_1^0\rangle=\langle H_2^0\rangle=v/\sqrt{2}$ 
\cite{BGH}. Three of the four degrees of freedom of each doublet are eaten 
by the longitudinal components of the $W^\pm$ and $Z^0$ gauge bosons of 
our world and the mirror world. Two scalar degrees of freedom remain as 
physical Higgses, with a squared mass matrix that reads
\be
\label{massmatrixTH}
\left(\begin{array}{cc}
2(\lambda+\gamma) v^2&  2\lambda v^2\\
2\lambda v^2 & 2 (\lambda+\gamma) v^2
\end{array}\right)\ ,
\ee
with eigenvalues
\bea
m_-^2&=&2\gamma v^2\ ,\nonumber\\
m_+^2&=&2(2\lambda+\gamma)v^2\ ,
\eea
and eigenvectors $h^0_\pm={\rm {Re}}(H_1^0\pm H^0_2)$. We see that the 
Higgs mass eigenstates are mixtures with 50$\%$ $H_1$ 
component and 50$\%$ $H_2$ component so that they have reduced couplings 
to matter and gauge bosons in our world. The eigenvector $h_+$  
corresponds to the  
Higgs excitations along the breaking direction and therefore, concerning 
the minimization condition, $m_+$ 
plays a role similar to $m_h$ in the SM (in fact, $m_+^2=-2\mu^2$).  
With $\gamma=0$, the global $SU(4)$ symmetry of the $(|H_1|^2+|H_2|^2)$ 
terms would result 
in a massless Goldstone boson in the direction transverse to the breaking, 
{\it i.e.} $m_-=0$. Having $\gamma\neq 0$ with $\gamma \ll \lambda$ avoids 
this and gives in a natural way a light Higgs in the spectrum \cite{TH}. 

Let us examine the structure of quadratically divergent corrections to the 
Higgs mass parameters in this model. As the $Z_2$ symmetry is not broken,
both $H_1$ and $H_2$ receive the same corrections, given by
\be
\label{quadTH}
\delta_q \mu^2 = {\Lambda^2\over 8\pi^2}\left[{3\over 
8}(3g^2+g'^2)+5\lambda+3\gamma-3\lambda_t^2\right]\ .
\ee
Notice that this formula assumes all the couplings in the mirror 
world take exactly the same values as in our world. 
We can write eq.~(\ref{quadTH}) in terms of particle masses by writing 
$5\lambda+3\gamma = (5m_+^2+m_-^2)/(4v^2)$. Then we find a result very 
similar to the SM case with the replacement $m_h^2\rightarrow 
(5m_+^2+m_-^2)/6$. If we fix $m_-$ to a low value, say $m_-=115$ GeV, the 
quadratic correction in (\ref{quadTH}) as a function of $m_+^2$ behaves 
a bit better than the SM quadratic correction as a function of $m_h^2$ 
\cite{BGH}.

A second difference with respect to the SM behaviour comes from a 
different RG evolution of the couplings in this case. We should compare 
the RGE for $(5\lambda + 3\gamma)$ in this model with that for 
$(3\lambda)$ in the SM, see eq.~(\ref{lambdaSM}) . Again we find a very 
similar result except for the replacement $4(3\lambda)^2\rightarrow 
(16/5)(5\lambda+3\gamma)^2
+4(3\gamma)^2$. For the case of interest, with $\gamma\ll \lambda$, we 
therefore conclude that RG effects in the quadratic 
corrections of this model are a bit softer than in the SM.  

Finally, as happened in the 2HDMs discussed before (see appendix~A), the 
constraints 
on 
the Higgs masses derived from EWPT are modified with respect to the SM 
ones. Now one has \cite{BGH}
\be
\label{EWPTw}
m_+ m_- < m_{EW}^2\ ,
\ee
where, as usual, $m_{EW}=\{186,219\}$ GeV is the EWPT indirect bound on 
the SM Higgs 
mass. 

\begin{figure}[t]
\vspace{1.cm} \centerline{
\psfig{figure=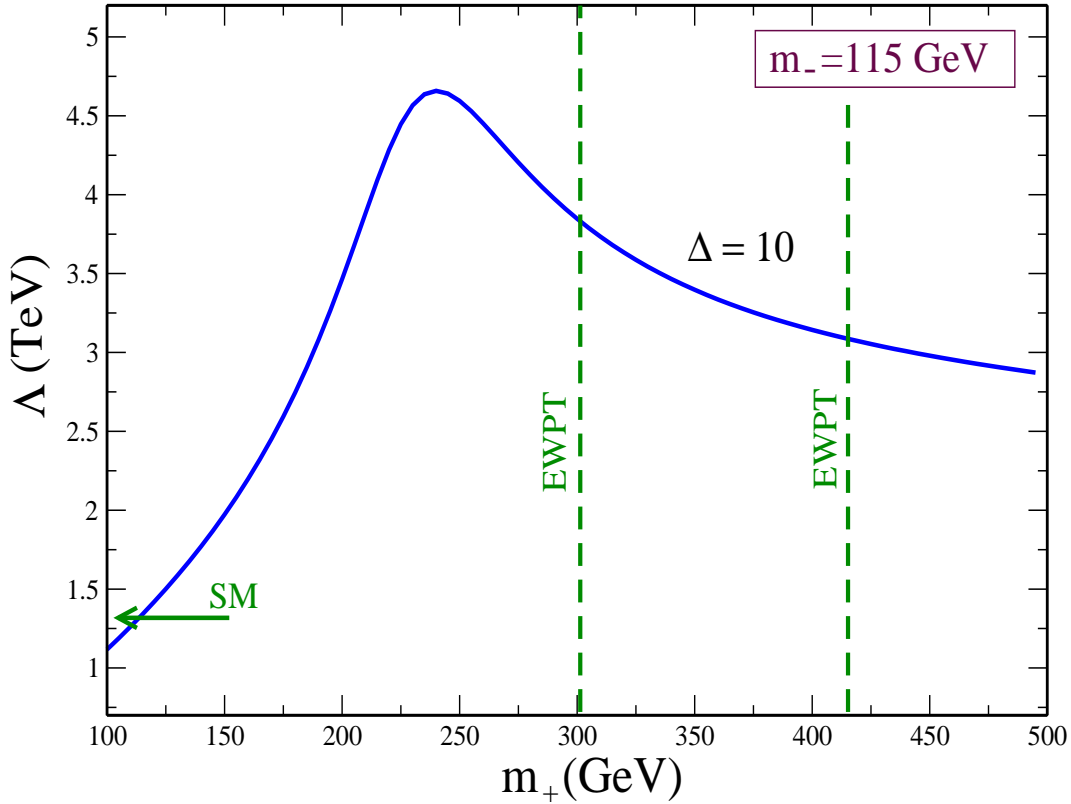,angle=-90,height=10cm,width=11cm,bbllx=4.cm,%
bblly=4.cm,bburx=20.cm,bbury=23.cm}
}
\caption{\footnotesize
Upper bound on the scale $\Lambda$ of New Physics from the requirement of 
less than 10\% tuning of EWSB in the Twin Higgs model with $m_-=115$ GeV. 
Shown by the dashed lines are the EWPT upper bounds on 
$m_+$. For comparison, the arrow marks the SM naturalness upper bound on 
$\Lambda$ for $m_h=m_-$.} 
\label{fig:TH0}
\end{figure}

As a result of the effects just discussed, this model is able to 
improve 
over the naturalness of the SM Higgs sector. Fig.~\ref{fig:TH0} shows 
the upper bound on the scale of New Physics, $\Lambda$ (we limit our 
analysis to a unique cut-off), that 
follows from imposing 
that the fine-tuning in $\mu^2$ (calculated as discussed 
in previous sections) is smaller than 10\% and choosing $m_-=115$ 
GeV. For comparison, the SM bound for $m_h=115$ GeV, {\it i.e.} $\Lambda< 
1.32$ TeV, 
is also indicated. However, it is more instructive to compare the 
bound on $\Lambda$ as a function of $m_+$ with the SM curve as a function 
of $m_h$, Figs.~\ref{fig:SM1} and~\ref{fig:SM3}. Then we see that the 
current curve has a 
shape very similar to the SM one, but it is a slightly bit higher. 
Moreover, the range of $m_+$ compatible with EWPT is also wider, fully 
including the maximum of the curve. Nevertheless, the improvement with 
respect to the SM situation is not dramatic.

Let us now discuss the case in which one introduces a small breaking 
of the $Z_2$ symmetry by considering different masses for $H_1$ and $H_2$. 
More explicitly, we add to the potential (\ref{VTH}) a term \cite{BGH}
\be
\delta V = m^2(|H_1|^2-|H_2|^2)\ .
\ee
With such modification, the minimum of the potential moves away from 
$\tan\theta\equiv \langle H_1^0\rangle/\langle 
H_2^0\rangle\equiv v_1/v_2=1$ 
(although we still have to keep $v_1=246$ GeV), with
\be
\cos 2\theta=-
\left({2\lambda+\gamma\over \gamma}\right){m^2\over \mu^2}
\ .
\ee 
The squared mass matrix 
for the two Higgses takes now the form
\be
\label{massmatrixTHm}
\left(\begin{array}{cc}
2(\lambda+\gamma) v_1^2&  2\lambda v_1 v_2\\
2\lambda v_1 v_2 & 2 (\lambda+\gamma) v_2^2
\end{array}\right)\ ,
\ee
with eigenvalues
\bea
m_-^2&\simeq &2\gamma (v_1^2c_\theta^2+v_2^2s_\theta^2)\ ,\nonumber\\
m_+^2&\simeq &2\lambda (v_1^2+v_2^2)+2\gamma 
(v_1^2s_\theta^2+v_2^2c_\theta^2)\ 
,
\eea
where we have expanded in $\gamma/\lambda$. The eigenvectors are 
defined as 
$h_-=\sqrt{2}{\rm Re}(c_\alpha H_1^0+s_\alpha H_2^0)$
and 
$h_+=\sqrt{2}{\rm Re}(-s_\alpha H_1^0+c_\alpha H_2^0)$. From 
(\ref{massmatrixTHm})
\be
\tan 2\alpha = {-\lambda\over\lambda+\gamma}\tan 2\theta\ .
\ee
For $\gamma\ll \lambda$, one has $\alpha\simeq -\theta$, so that $h_+$ is 
still aligned with the breaking direction and its mass is still of direct 
relevance for the naturalness of electroweak breaking (again $m_+^2\simeq 
-2\mu^2$).

\begin{figure}[t]
\vspace{1.cm} \centerline{
\psfig{figure=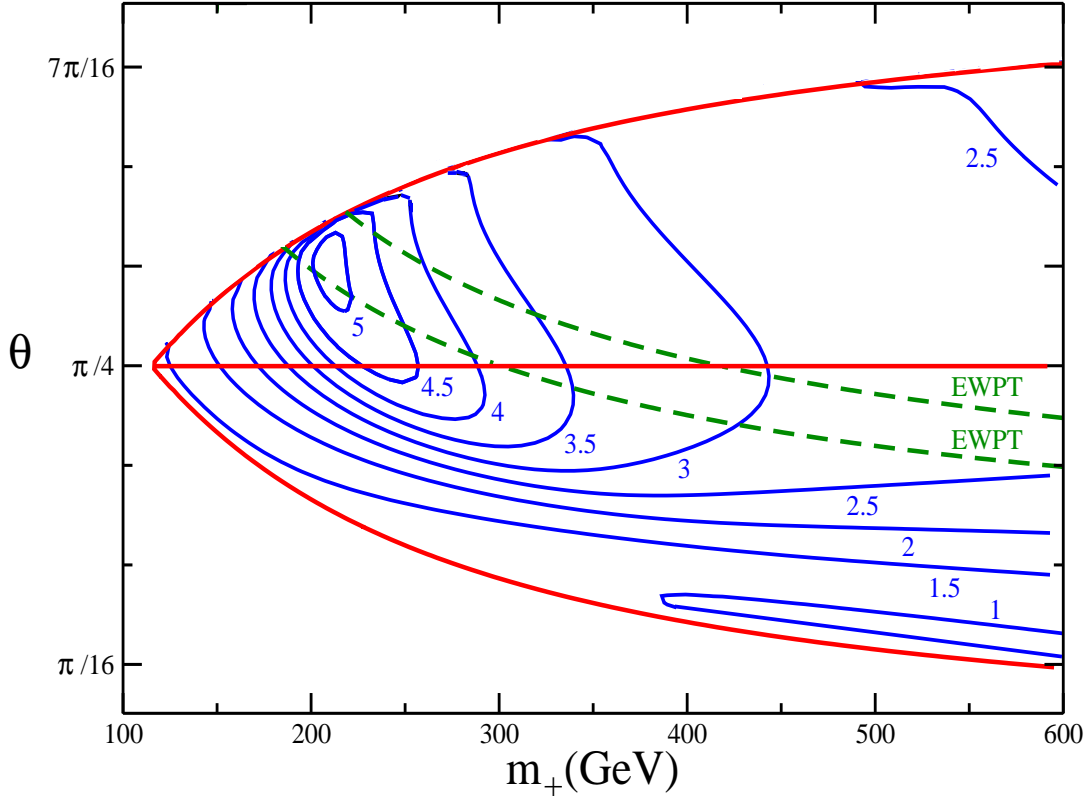,angle=-90,height=10cm,width=11cm,bbllx=4.cm,%
bblly=4.cm,bburx=20.cm,bbury=23.cm}
}
\caption{\footnotesize Contour lines of the 10\% naturalness upper bounds 
on the scale of New Physics $\Lambda$ (in TeV) in the Twin Higgs model 
with $m\neq 0$ and $m_-=115$ GeV. The dashed lines show the EWPT upper 
bounds on $m_+$.}
\label{fig:pez}
\end{figure}

Before presenting the results for the fine-tuning in the case $m\neq 0$, 
notice that $\lambda$ and $\gamma$ in (\ref{massmatrixTHm}) can be 
obtained in terms of $m_+$ and 
$m_-$ as
\bea
\label{ldm}
\lambda& =& \pm {1\over 2v_1v_2}\sqrt{(m_+^2s_\theta^2-m_-^2c_\theta^2)
(m_+^2c_\theta^2-m_-^2s_\theta^2)}\nonumber\ ,\\
\gamma&=&{m_+^2+m_-^2\over 2(v_1^2+v_2^2)}-\lambda\ .
\eea
It follows that, for fixed 
$m_+$, $m_-$ and $\theta$, there are two different solutions for $\lambda$ 
and $\gamma$ with different signs for $\lambda$ (the region $\lambda<0$ is 
accessible provided $|\lambda|<\gamma/2$, to avoid an instability in the 
scalar potential). It can be shown that the best case for naturalness 
corresponds to $\lambda>0$ and we restrict our analysis to that case.

We can also see from eqs.~(\ref{ldm}) that the parameter space is 
limited to the region $m_+\geq m_- {\rm Max}
\{\tan\theta, 1/\tan\theta\}$. This is 
shown in Fig.~\ref{fig:pez} which corresponds to the case $m_-=115$ GeV: 
the accessible parameter space lies inside the ``fish'' profile. 
For any $\theta$, the minimal 
value of $m_+$ corresponds to taking 
$\lambda\rightarrow 0$ in the mass matrix (\ref{massmatrixTHm}). In that 
limit, the mass matrix is diagonal,  
$H_1$ is the SM Higgs with mass $m_h^2=2\gamma v_1^2$ and $H_2$ is the 
Higgs boson of the mirror world, with mass $m_{h'}^2=2\gamma v_2^2$. In 
Fig.~\ref{fig:pez}, this limit corresponds to the boundary of 
the 
allowed region of parameter space. Along the upper limit, with 
$\tan\theta>1$ one has $m_+\equiv m_h\geq m_-\equiv m_{h'}$. Therefore, 
$m_+$ plays the role of the mass of the SM Higgs boson along 
that line.
For the lower limit of parameter space in Fig.~\ref{fig:pez}, with 
$\tan\theta<1$, one has 
instead $m_+\equiv m_{h'}\geq m_-\equiv m_{h}$ and therefore, along that 
line 
$m_+$ is simply the mass of the mirror Higgs, totally decoupled from our 
world (which has a Higgs mass fixed to $m_h=115$ GeV). We have also marked 
in Fig.~\ref{fig:pez} the line $\theta=\pi/4$, which corresponds to 
$m=0$.

The comments above are very useful to understand the behaviour of the 
fine-tuning associated to EWSB in the general
case with $m\neq 0$. Before discussing them, let us remark that, in the 
case with $v_1\neq v_2$ we are really interested in the tuning associated 
with electroweak breaking in our world, and therefore in the tuning 
necessary to get right $v_1$, which is fixed by the minimization condition
\be
v_1^2=-{\mu^2\over 2\lambda+\gamma}-{m^2\over\gamma}\ .
\ee
The upper bounds on $\Lambda$ coming from requiring less than $10\%$ 
tuning are shown by Fig.~\ref{fig:pez}
as contour lines (in TeV)  in the parameter space 
$\{m_+,\theta\}$ for $m_-=115$ GeV. We can recognize the SM numbers (see 
Fig.~\ref{fig:SM1} or \ref{fig:SM3}) along 
the upper limit of the allowed region of parameter space, where 
$m_+$ is precisely the SM Higgs mass. Along the line $\theta=\pi/4$ we can 
recognize the numbers corresponding to the $m=0$ case shown in 
Fig.~\ref{fig:TH0}. Finally, along the lower limit of the allowed 
parameter space we recover the upper bound $\Lambda\simeq 1.3$ TeV, 
corresponding to the SM case with $m_-=115$ GeV.
The plot also shows the constraint on $m_+$ and $m_-$ from 
EWPT (see appendix~A), which reads \cite{BGH}
\be
\label{EWPTm}
m_+^2< m_{EW}^2\left[{m_{EW}^2\over m^2_-}\right]^{1/\tan^2\alpha}\ .
\ee
Equation (\ref{EWPTm}) generalizes 
(\ref{EWPTw}) for the case of $m\neq 0$. The region above these lines 
(corresponding to the two cases $m_{EW}=186$ GeV and 219 GeV) is 
disfavored.  From this plot we conclude again that, even though the 
upper 
bound on the scale of New Physics can be higher than in the SM (the best 
point, giving $\Lambda\simgt 5$ TeV, corresponds to $m_+\sim 215$ GeV and 
$\theta\sim 5\pi/16$), the global 
effect is never dramatic.

\section{The SM and the Hierarchy Problem Revisited}

We have learned from the previous examples that it is not easy to improve
significantly the naturalness of the SM by complicating the Higgs sector. 
Such complication
could be justified, however, to explain the hypothetical detection at LHC
of a heavy Higgs or the non-detection of strongly coupled NP 
able to cancel the top quadratic contributions to the Higgs mass.
In this section we take a different view and explore the possibilities
of the pure SM, as effective theory, to accommodate such possibilities.

First, let us briefly consider the case of a heavy SM Higgs, i.e.
well above the range allowed from EWPT (which require $m_h<186-219$ GeV
at 95\% c.l.). As is well known, a heavy Higgs would conflict 
mainly with the measured value of the $T$ parameter, see eq.~(\ref{T1}), 
so that new physics beyond the SM is needed to reconcile theory and 
experiment.
In the IDM \cite{BHR} discussed in subsect.~3.1 such extra physics comes
from the extension of the Higgs sector (to a particular 2HDM). However,
the same NP that cancels the SM quadratic contributions to the Higgs mass 
parameter could fix $\Delta T$ \cite{BHR}. 
E.g. it is well known that a non-renormalizable operator
\be
\label{HOO}
{\kappa_6^2\over \Lambda^2}\left|H^\dagger D_\mu H\right|^2\ ,
\ee
where $\kappa_6^2$ is a dimensionless coefficient and $\Lambda$ is the 
scale 
of NP, gives a contribution to $T$ 
\be
\Delta T \simeq -{\kappa_6^2\over 2\alpha_{\rm em}}
{v^2\over \Lambda^2}\ 
\simeq\  - \kappa_6^2\left({2\ {\rm TeV}\over \Lambda}\right)^2\ ,
\ee
that could move the final value of $T$ inside the experimental range.
E.g. for $m_h=400$ GeV, this requires $\kappa_6^2>0$ and $\Lambda$  
in the range $\kappa_6\times 4$ TeV, 
something perfectly consistent with the 
naturalness upper bounds on $\Lambda$ (see 
solid lines for $\Delta=10$ in Figs.~\ref{fig:SM1}--\ref{fig:SM3} of 
sect.~2).
In that case, no hint of NP, apart from higher order operators, 
would be left below $\Lambda$, e.g. the spectrum of the
effective theory would be just the SM one, with no modified Higgs sector.

Let us now consider the possibility that LHC does not find any NP able
to cancel the top (or the other) quadratic contributions to the 
Higgs mass parameter,
or perhaps the most extreme (and unpleasant) situation that, 
apart from the Higgs itself, no NP is found at all.
In particular we want to know if that situation
could be consistent (with no fine-tunings) with just the SM as the 
effective theory.

Assuming a unique cut-off for all the quadratically divergent loops, the 
fine-tuning status of the SM was shown in Fig.~\ref{fig:SM1} (upper solid 
line). Actually, once the Higgs mass is measured one should not include 
$\Delta_\lambda$ in the fine-tuning analysis. In that future time, the 
curve describing the fine-tuning will be the black dashed line of 
Fig.~\ref{fig:SM1}, which shows a deep throat around $m_h=225$ GeV. So, 
{\em if} in the future the Higgs happens to have a mass in the region 
$\sim 225\pm 25$ GeV, the NP could easily escape detection with no 
fine-tuning price. Such change of conclusion might seem paradoxical (why 
should we wait until a future measurement for such an statement?), but 
indeed is not strange at all and is characteristic of fine-tuning 
arguments: being based on statistical considerations, the conclusions may 
vary according to our partial (and time-dependent) knowledge of the 
relevant parameters in the problem. At present, however, we should 
establish the fine-tuning estimates using our actual knowledge of the 
physical parameters, so the upper solid line of Fig.~\ref{fig:SM1} gives 
the present status. Still we see that the cut-off could lie as far as 3-4 
TeV for $m_h$ in the $200$ GeV region, which might allow the NP to escape 
LHC detection (depending on what NP it is). A Higgs in that region would 
be still consistent with EWPT (even if marginally), so this is certainly a 
possibility to keep in mind. Moreover, as discussed above, higher order 
operators like (\ref{HOO}) can improve the consistency with EWPT.

\begin{figure}[t]
\vspace{1.cm} \centerline{
\psfig{figure=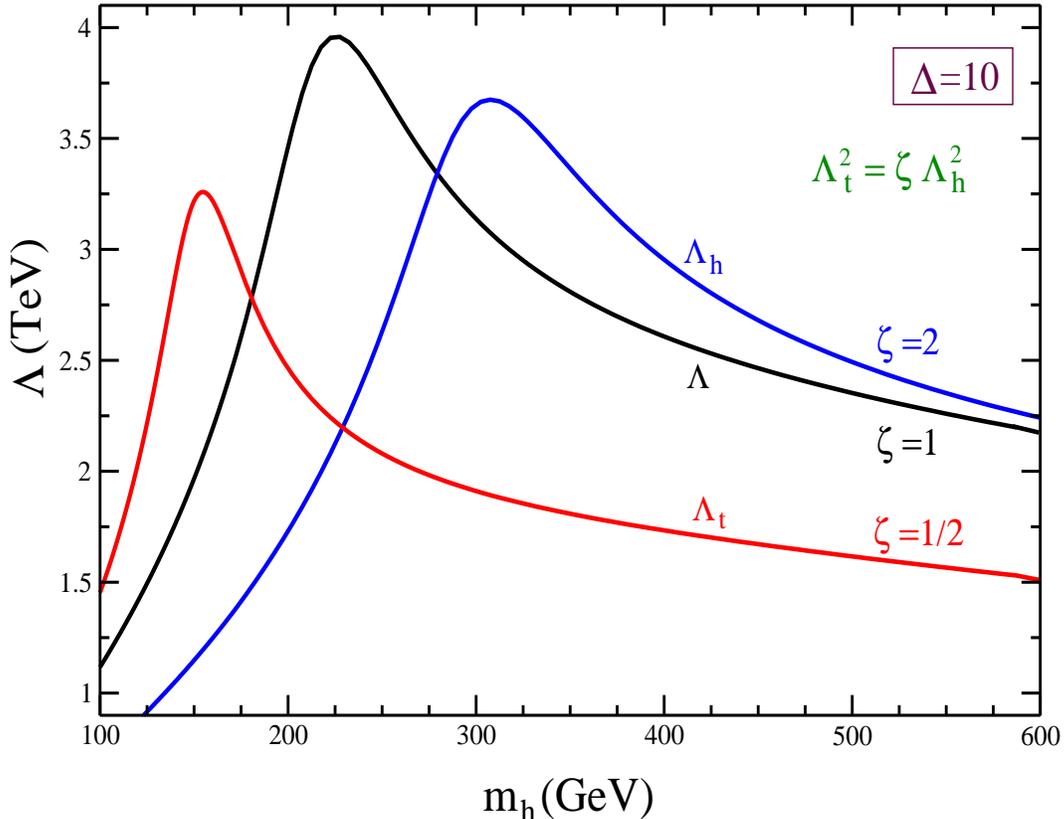,angle=-90,height=10cm,width=11cm,bbllx=4.cm,%
bblly=4.cm,bburx=20.cm,bbury=23.cm}
}
\caption{\footnotesize SM naturalness upper bound on the scale of NP for 
$\Delta=10$ with cut-offs correlated as $\Lambda_t^2=\zeta \Lambda^2_h$ 
for different values of $\zeta$.} 
\label{fig:zetaLambda}
\end{figure}

However, we could face in the future the situation that a light Higgs,
say $m_h\simeq 150$ GeV, is discovered at the LHC, but no other signal of 
NP is 
found. Would this necessarily represent a fine-tuned situation? Probably, 
the most pleasing answer would be ``yes'', since it would imply that this 
undesirable
situation is unlikely. Nevertheless, the answer could be different.
As discussed in sect.~2, it is perfectly possible that the
cut-offs of all quadratically divergent contributions are
correlated, namely when they arise from the same piece of NP (as can 
occur when the NP is SUSY). But notice that this
does not mean that all cut-offs should be exactly equal. Again SUSY is a 
good 
example: squark masses
(in particular stop masses) do not need to be equal to the Higgsino
masses: the former are determined by the high energy soft masses
plus RG effects, while the second depend on the $\mu$ parameter
(which can be correlated with, but is not necessarily equal to, $m_{\rm 
soft}$).
Therefore it is likely that $\Lambda_t$ and $\Lambda_h$
are correlated but not equal, say $\Lambda_t^2 = \zeta \Lambda_h^2$
with some proportionality factor $\zeta={\cal O}(1)$, which depends
on the unknown NP. This should be plugged in eq.~(\ref{quadrdiv0})
to re-evaluate $\delta_q m^2$ and the fine-tuning $\Delta$.
Obviously, varying $\zeta$, even slightly, shifts the value
of $\lambda$ (and thus of $m_h$) where the approximate cancellation 
of the quadratic contributions takes place. Consequently, the
position of  Veltman's throat changes, as shown in 
Fig.~\ref{fig:zetaLambda},
where the cut-off plotted is always the smallest one, {\it i.e.} 
${\mathrm Min}\ \{\Lambda_h , \Lambda_t\}$.
Interestingly, taking a modest $\zeta = 1/2$ the throat is around 
$m_h=150$ GeV,
with $\Lambda\sim 3$ TeV. In this situation NP could escape
LHC detection with no fine-tuning.
It is remarkable how correlated, but slightly different, 
cut-offs change in a physically significant 
way the ordinary expectations about the (approximate)
cancellation of the quadratically divergent contributions. This 
can be crucial for the detection or non-detection of NP at LHC.
As argued above, this situation is perfectly possible and 
makes less compelling the ordinary hierarchy-problem argument 
to expect NP showing up at LHC. This is also in line with the
conclusions of ref.~\cite{CEHI} in the sense that fine-tuning arguments
can only be applied reliably when a concrete example of NP is assumed.
Although the general bound of eq.~(\ref{quadrft}) is usually valid
(and even conservative),
${\cal O}(1)$ factors in that equation are crucial for the visibility 
of NP at LHC.

One may also wonder about the validity of using the SM as the effective
theory at scales between the smallest and the largest cut-off. E.g. if
$\Lambda_h>\Lambda_t$, how reliable is it to evaluate Higgs-loops
within the SM above $\Lambda_t$? This question was already raised
in the context of the BH model, and briefly discussed in footnote 3.
Of course, the answer depends on the particular sort of NP entering at 
$\Lambda_t$. In any case, this kind of analyses at least indicates
that the dangerous radiative
corrections may (approximately) cancel for different values of $m_h$,
depending of the characteristics of the NP. 

We can make a more precise
statement by using the approach of ref.~\cite{CEHI}: 
modelling the NP as a 
set of new particles with (possibly) $h$-dependent 
masses, the cancellation of the SM
quadratically divergent corrections requires
\be
\left.{\partial^2{\mathrm Str} M^2\over \partial h^2}\right|_{h=0}  
\equiv \sum_a^{\small \rm SM}N_a  \left.{\partial^2 m_a^2\over \partial 
h^2}
\right|_{h=0}  + \sum_b^{\small \rm NP}N_b \left.{\partial^2 
m_b^2\over \partial h^2}
\right|_{h=0}=0\ ,
\ee 
where $m_a, N_a$ ($m_b, N_b$) are the masses and multiplicities 
(with negative sign for fermions) of the SM (NP) states. Then,
the logarithmic and finite
contributions of NP to $m^2$ are given, 
in the $\overline{\rm MS}$ scheme,
by
\be
\label{deltam1MS}
\delta_{\rm NP}^{\overline{\rm MS}}\ m^2 =
 \sum_b^{\small \rm NP}
{N_b\over 32 \pi^2}
\left[ {\partial^2 
m_b^2\over \partial h^2}
m_b^2\left( \log {m_b^2\over Q^2}-1\right)+
\left({\partial m_b^2\over \partial h}\right)^2
\log {m_b^2\over Q^2} \right]_{h=0}\ ,
\ee
where $Q$ is the renormalization scale, to be identified with a
high-energy cut-off scale, $\Lambda_{\mathrm HE}$, which sets the limit 
of validity of the NP description (it could be as large as $M_{Pl}$). 
Quantitatively, these contributions are roughly similar to the 
SM quadratically-divergent one, eq.~(\ref{quadrdiv0}), replacing 
$m_b\rightarrow \Lambda$. Hence, the naive estimate (\ref{quadrdiv0})
is reasonable [and even conservative due to the 
logarithmically-enhanced terms in (\ref{deltam1MS})], but we 
can expect different
"cut-offs" associated to the different $m_b$ masses, couplings and RG 
effects in (\ref{deltam1MS}).

\section{Summary and Conclusions}

Generic arguments based on the size of quadratically-divergent 
contributions to the Higgs mass (Big Hierarchy problem) imply the 
existence of New Physics (NP) beyond the SM at a scale $\Lambda\simlt$ few 
TeV. The precise size of $\Lambda$ depends on the degree of fine-tuning 
one is willing to tolerate: for 10\% fine-tuning and a Higgs mass 
$m_h=115-200$ GeV, then $\Lambda\simlt 2-3$ TeV, hopefully within the 
reach of LHC.

There are some reasons to consider possible departures from
this simple SM scenario. 

{\em a)} The above upper bound on $\Lambda$ is generically in some 
tension with the experimental lower bounds on the suppression scale of 
higher order operators, derived from electroweak precision tests (EWPT), 
which typically require $\Lambda\simgt$ 10 TeV (Little 
Hierarchy problem). This tension would be relaxed if the upper bound on 
$\Lambda$ could be pushed up to the ${\cal O}(10)$ TeV region, which
would imply that the NP responsible for the cancellation of the 
quadratic divergences would escape LHC detection. Actually, one should be 
prepared to face the possibility (admittedly unpleasant) that no NP apart 
from the Higgs is found at LHC, in spite of naturalness arguments.

{\em b)} It could happen
that the Higgs found at LHC is beyond the range consistent with EWPT
($m_h\simlt 186-219$ GeV).

Both cases may be interpreted as pointing out to some 
departure from the ordinary SM Higgs sector. One of the simplest 
and best motivated
modifications of the SM Higgs sector one can think
of is the two Higgs doublet model (2HDM). Several 
recent works \cite{BGH,BH,BHR} have examined  the 
capabilities of such scenarios to address the 
previous questions. 
These 2HDMs include: the ``Inert Doublet Model'' (IDM), 
the ``BH Model'' and the Twin-Higgs
Model. The conclusions of \cite{BGH,BH,BHR} are that these models can 
raise
the scale of new physics above the LHC reach (keeping naturalness
under control) with a light or heavy Higgs (depending on the model), 
in a way consistent with EWPT. Then the NP responsible for the cancellation
of the SM quadratic divergences could escape LHC, although we
could observe a modified Higgs sector. Thus these models are claimed
to ``improve the naturalness'' of the pure SM.

Our first goal in this paper has been to perform a careful and fair comparison 
of the naturalness of these modified-Higgs-sector models with that of the SM. 
This requires a sensible criterion to quantify 
the degree of fine-tuning, which should be applied to all the models 
and computed as rigorously as possible (including e.g. 
radiative effects). Our conclusions are that, generically, these 
models do not improve the naturalness of the SM, i.e. they are not
able to push significantly the NP cut-off beyond the SM estimate. 
Actually, in many cases the naturalness is  worsened rather than
improved. This is in part because
the structure of the models normally requires additional fine-tunings
of a size similar to that required for a correct EW 
breaking.
This result is common in models with a structure more 
complicated than that of the SM. Normally, epicycles are penalized in 
naturalness estimates (which is somehow satisfactory).

This does not mean that these models are not interesting. They are 
well motivated and show how a LHC phenomenology different from
the pure SM expectations could take place. E.g. the IDM shows explicitly
how a heavy Higgs can be accommodated in a way consistent with EWPT.
On the other hand, not all the models are on the same footing with respect 
to naturalness. The Twin-Higgs model behaves the best; in fact, it is 
the only one
able to improve the SM naturalness, though not dramatically.

Our second goal in the paper has been to examine if the SM {\em alone}
could be able to cope with the above-mentioned ``unexpected'' situations 
at LHC:
i)  a heavy SM Higgs
(well above the range allowed from EWPT, $m_h<186-219$ GeV
at 95\% c.l.) and/or ii) 
the possibility that LHC does not find any NP able
to cancel the quadratic contributions to the 
Higgs mass,
or perhaps the most extreme (and unpleasant) situation that, 
apart from the Higgs itself, no NP is found at all.

Regarding question i), it is well-known that a heavy Higgs is especially 
harmful for the $T$--parameter. However, these dangerous contributions to 
$T$ could be compensated by the same NP which is responsible for the 
cancellation of the quadratic contributions to $m_h$. I.e. there is no 
need to modify the Higgs sector of the effective theory below the scale of 
NP (as the IDM does), although of course that is a possibility. We have 
emphasized that higher dimension operators suppressed by a scale $\Lambda$ 
consistent with the naturalness bound can do the job. In that case, no 
hint of NP, apart from higher order operators, would be left below 
$\Lambda$, e.g. the spectrum of the effective theory would be just the SM 
one, with no modified Higgs sector.

Regarding question ii) we have shown first that if $m_h$ lies
in the $200$ GeV region, $\Lambda$ can be as large as 3-4 TeV,
 which might allow the NP to escape LHC detection (depending on what NP it 
is). This is because at $m_h\sim 225$ GeV there is an accidental 
cancellation between the top and Higgs loop-contributions to the Higgs 
mass (assuming the same cut-off for both kinds of loops, i.e. 
$\Lambda_t=\Lambda_h$). Consequently the fine-tuning is suppressed and 
$\Lambda$ can be larger (this is the so-called ``Veltman's throat'', 
visible in 
Fig.~\ref{fig:SM1}). A Higgs in that region would still be consistent with 
EWPT (even if marginally), and so this is certainly a possibility. 
Moreover, as mentioned above, higher order operators can improve the 
consistency with EWPT.

If the Higgs is lighter, say $m_h\simlt 150$ GeV, the cancellation is not 
efficient and $\Lambda$ falls below 2 TeV, probably within LHC reach 
(see 
again Fig.~\ref{fig:SM1}). However, this prospect changes if the 
different cut-offs are correlated but not exactly equal. Parametrizing 
$\Lambda_t^2 = \zeta \Lambda_h^2$ (where $\zeta$ depends on the kind of 
NP), we have shown that varying $\zeta$, even slightly, shifts the value 
of $m_h$ where the approximate cancellation of the quadratic contributions 
takes place. Consequently, the position of Veltman's throat changes, as 
shown in Fig.~\ref{fig:zetaLambda}. E.g. taking a modest $\zeta = 1/2$ the 
throat is around $m_h=150$ GeV, with $\Lambda\sim 3$ TeV. In this 
situation NP could escape LHC detection with no fine-tuning. This makes 
less compelling the ordinary hierarchy-problem argument to expect NP 
showing up at LHC. Although the general bound of eq.~(\ref{quadrft}) is 
usually valid (and even conservative), ${\cal O}(1)$ factors in that 
equation (e.g. arising from taking $\Lambda_t \neq \Lambda_h$, as 
mentioned) are crucial for the visibility of NP at LHC.

To conclude, it is not easy to improve the naturalness of the SM by 
modifying the Higgs sector. Moreover, although the general arguments give good
prospects to expect NP within the reach of LHC, it may anyway happen that 
NP escapes LHC detection without a fine-tuning price while leaving the 
pure SM as the sole effective theory valid at LHC energies.

\section*{Appendix A}
\setcounter{equation}{0}
\renewcommand{\theequation}{A.\arabic{equation}}

In this Appendix we recall the well known expression \cite{T-THDM} for the 
$\Delta T$
contribution from two Higgs doublets and then particularize it to the 
three different scenarios discussed in section~3. The rotation angle 
$\alpha$ between the $CP$-even neutral Higgs fields $h_1$ and $h_2$ and 
the mass eigenstates $h_-$ and $h_+$ (with $m_-\leq m_+$) is defined by 
\be
\left[\begin{array}{c}
h_-\\
h_+
\end{array}
\right]=\left(
\begin{array}{cc}
\cos\alpha &\sin\alpha\\
-\sin\alpha & \cos\alpha
\end{array}
\right)\left[\begin{array}{c}
h_1\\
h_2
\end{array}
\right]\ ,
\ee
while $\beta$ is simply given by $\tan\beta=v_2/v_1$ as usual. The 
Higgs contribution to the $T$ parameter can be conveniently written as the 
sum of two different pieces, one SM-like contribution related to 
excitations along the breaking direction, $h_\parallel\equiv h_1\cos\beta 
+h_2\sin\beta$, and one additional correction related to excitations 
along the transverse direction $h_\perp\equiv -h_1\sin\beta + 
h_2\cos\beta$.
The explicit expressions are as follows. For the parallel contribution
\be
\Delta T_\parallel = \sin^2(\alpha-\beta)\Delta 
T_{SM}(m_+)+\cos^2(\alpha-\beta)\Delta T_{SM}(m_-)\ ,
\label{Tpar}
\ee
where 
\be
\Delta T_{SM}(m)={1\over 4\pi s_W^2M_W^2}\left[A(m,M_W)+
{M_W^2m^2\over m^2-M_W^2}\log{m^2\over M_W^2}-(M_W\rightarrow M_Z)
\right]\ ,
\ee
and 
\be
A(x,y)\equiv {1\over 8}\left(x^2+y^2-{2x^2y^2\over x^2-y^2}\log{x^2\over 
y^2}
\right)\ .
\ee
For later use we also recall that, for not very hierarchical $x/y$, one 
can use the approximation \cite{BHR}
\be
\label{ap}
A(x,y)\simeq {1\over 6} (x-y)^2\ .
\ee
In addition, for large $m$ one has 
\be
\Delta_{SM}(m)\simeq {\rm Const.} - {3\over 8 \pi \cos^2 
\theta_w}\ln{m\over M_Z}\ ,
\ee
so that the EWPT constraint on $\Delta T$ is a constraint on $\log m$.

The transverse correction is
\bea
\Delta T_\perp &=& {1\over 4\pi s_W^2M_W^2}\left\{A(m_A,m_{H^+})+
\sin^2(\alpha-\beta)\left[A(m_{H^+},m_-)-A(m_A,m_-)\right]\right.\nonumber\\
&&\left.\hspace{2cm}+
\cos^2(\alpha-\beta)\left[A(m_{H^+},m_+)-A(m_A,m_+)\right]
\right\}\ ,
\label{Tper}§
\eea
where $m_A$ is the mass of the pseudoscalar Higgs and $m_{H^+}$ the mass 
of 
the charged Higgs.

\subsection*{A.1 Inert Doublet Model}

In this model one has $\beta=0$ and $\alpha=0$ (when $m_-=m_{h_1}$) or 
$\alpha=\pi/4$ (when $m_+=m_{h_1}$). To treat both cases simultaneously 
call $m_h$ the mass along $h_1$ and $m_S$ the mass along $h_2$. We 
then get 
\bea
\Delta T_\parallel &=& \Delta T_{SM}(m_h)\ ,\nonumber\\
\Delta T_\perp &=&  {1\over 4\pi s_W^2M_W^2}\left[ 
A(m_A,m_{H^+})+A(m_{H^+},m_S)-A(m_A,m_S)\right]\ ,
\eea
and using (\ref{ap})
\be
\Delta T_\perp \simeq  {1\over 12 \pi s_W^2 
M_W^2}(m_{H^+}-m_A)(m_{H^+}-m_S)\ .
\ee

\subsection*{A.2 Barbieri-Hall Model}

In the particular case with small mixing angle $\alpha$ one gets
\bea
\Delta T_\parallel &=& \sin^2\beta\ \Delta T_{SM}(m_+)+\cos^2\beta\ \Delta 
T_{SM}(m_-)\ ,\nonumber\\
\Delta T_\perp &=& {1\over 4\pi s_W^2M_W^2}\left[A(m_A,m_{H^+})+
\sin^2\beta\left[A(m_{H^+},m_-)-A(m_A,m_-)\right]\right.\nonumber\\
&&\left.\hspace{2cm}+
\cos^2\beta\left[A(m_{H^+},m_+)-A(m_A,m_+)\right]
\right\}\ .
\eea
For $m_A\simeq m_{H^+}$ one can neglect $\Delta T_\perp$. In that case the 
EWPT bound on $\Delta T$ is a constraint on $\sin^2\beta\ 
\log m_+ +\cos^2\beta\ \log m_-$.

If $\alpha$ is not small, one should use the general formulae 
(\ref{Tpar}) and (\ref{Tper}). It is still true that $\Delta T_\perp$ is 
negligible for $m_A\simeq m_{H^+}$.

\subsection*{A.3 Twin Higgs}

In this case only $H_1$ is a $SU(2)_L$ Higgs doublet but one can still use 
the 
general formulas (\ref{Tpar}) and (\ref{Tper}) simply setting $\beta=0$ 
and $m_A=m_{H^+}=0$ ($A^0$ and $H^+$ would in fact be Goldstone bosons 
coming from $H_2$). One then gets
\bea
\Delta T_\parallel &=& \sin^2\alpha\ \Delta 
T_{SM}(m_+)+\cos^2\alpha\ \Delta
T_{SM}(m_-)\ ,\nonumber\\
\Delta T_\perp &=& 0\ ,
\eea
and EWPT constrain the combination $\sin^2\alpha\ \log m_+ +\cos^2\alpha\ 
\log m_-$.

\vspace{0.3cm}
\noindent
{\large {\bf Acknowledgments}} We thank Z. Chacko and A. Romanino
for useful discussions and clarifications.
This work is supported in part by the Spanish
Ministry of Education and Science, through a M.E.C. project (FPA2001-1806)
and by a Comunidad de Madrid project (HEPHACOS; P-ESP-00346).
The work of Irene Hidalgo has been supported by a FPU grant from the 
M.E.C.


\end{document}